\documentclass[
aps,%
prb,%
amsmath,%
amssymb,%
reprint,%
superscriptaddress,%
citeautoscript,%
floatfix%
]{revtex4-2}

\usepackage[utf8]{inputenc}
\usepackage{graphicx}
\usepackage{dcolumn}
\usepackage{bm}
\usepackage{amssymb}
\usepackage{amsmath}
\usepackage{graphicx}
\usepackage{color}
\usepackage{xcolor}
\usepackage[colorlinks,linkcolor=blue,citecolor=blue,urlcolor=blue]{hyperref}
\usepackage{xparse}
\usepackage{enumitem}
\usepackage{xkvltxp}
\usepackage{xcolor}
\usepackage{comment}
\usepackage{ifthen}
\usepackage{siunitx}

\newcommand{\Jel}{\hat{J}_{\mathrm{el}}}
\newcommand{\Jph}{\hat{J}_{\mathrm{ph}}}
\newcommand{\LL}{\hat{L}}
\newcommand{\R}{\mathbf{R}}
\newcommand{\Q}{\mathbf{Q}}
\newcommand{\Hop}{\hat{\mathcal{H}}}
\newcommand{\F}{{\mathbf{F}}}
\newcommand{\dipole}{\hat{\boldsymbol{\mu}}}
\newcommand{\C}[1]{\hat{C}_{#1}}
\newcommand{\asym}{$a_1^1e^3$}
\newcommand{\sym}{$a_1e_x^{1.5}e_y^{1.5}$}

\newcommand{\ket}[1]{
  \mathchoice%
  {\left|#1\right\rangle}       
  {|#1\rangle}                  
  {|#1\rangle}                  
  {|#1\rangle}                  
}
\newcommand{\braket}[2]{\left\langle#1\middle|#2\right\rangle}
\newcommand{\matrixel}[3]{\left\langle#1\middle|#2\middle|#3\right\rangle}
\newcommand{\tE}{^{3}\!E}
\def\tA2{^{3}\!A_2}
\def\sA1{^{1}\!A_1}
\newcommand{\sE}{^{1}\!E}

\NewDocumentCommand{\gwf}{s O{i} O{n}}{
  \IfBooleanTF#1
  {\Psi_{#3}^{#2}}
  {\ket{\Psi_{#3}^{#2}}}
}
\NewDocumentCommand{\ggwf}{s O{i} O{n}}{
  \IfBooleanTF#1
  {\Phi_{#3}^{#2}}
  {\ket{\Phi_{#3}^{#2}}}
}
\NewDocumentCommand{\wf}{O{i}}{\ket{\Phi_{#1}}}
\NewDocumentCommand{\vwf}{s O{i} O{n} O{} O{}}{
  \chi_{#2#3}^{#4}
  \IfBooleanTF#1
  {}
  {(\Q_{#5})}
}

\newboolean{draft}   
\setboolean{draft}{true}
\NewDocumentCommand\todo{m}%
{%
  {\color{blue} (TODO: #1)}%
}
\NewDocumentCommand\change{om}%
{%
  \ifthenelse{\boolean{draft}}{
    \IfNoValueTF{#1}{}%
    {%
      {\color{gray}[#1]}
    }%
    {\color{orange}#2}%
  }%
  {#2}%
}

\begin{document}


\title{Vibrational and vibronic structure of isolated point
  defects:\texorpdfstring{\\}{ } the nitrogen--vacancy center in
  diamond}

\author{Lukas Razinkovas}%
\email{lukas.razinkovas@ftmc.lt}%
\affiliation{Center for Physical Sciences and Technology (FTMC),
  Vilnius LT-10257, Lithuania}

\author{Marcus W. Doherty}%
\affiliation{Laser Physics Centre, Research School of Physics,
  Australian National University, Canberra, Australian Capital
  Territory 2601, Australia}

\author{Neil B. Manson}%
\affiliation{Laser Physics Centre, Research School of Physics,
  Australian National University, Canberra, Australian Capital
  Territory 2601, Australia}

\author{Chris G. Van de Walle}%
\affiliation{Materials Department, University of California, Santa
  Barbara, California 93106-5050, USA}

\author{Audrius Alkauskas}%
\affiliation{Center for Physical Sciences and Technology (FTMC),
  Vilnius LT-10257, Lithuania}

\date{\today}


\begin{abstract}
  We present a theoretical study of vibrational and vibronic
  properties of a point defect in the dilute limit by means of
  first-principles density functional theory calculations. As an
  exemplar we choose the negatively charged nitrogen--vacancy center, a
  solid-state system that has served as a testbed for many protocols
  of quantum technology. We achieve low effective concentrations of
  defects by constructing dynamical matrices of large supercells
  containing tens of thousands of atoms. The main goal of the paper is
  to calculate luminescence and absorption lineshapes due to coupling
  to vibrational degrees of freedom. The coupling to symmetric $a_1$
  modes is computed via the Huang--Rhys theory. Importantly, to
  include a nontrivial contribution of $e$ modes we develop an
  effective methodology to solve the multi-mode $E \otimes e$
  Jahn--Teller problem. Our results show that for NV centers in diamond
  a proper treatment of $e$ modes is particularly important for
  absorption. We obtain good agreement with experiment for both
  luminescence and absorption. Finally, the remaining shortcomings of
  the theoretical approach are critically reviewed. The presented
  theoretical approach will benefit identification and future studies
  of point defects in solids.
\end{abstract}


\maketitle


\section{Introduction\label{sec:intro}}

The calculation of optical lineshapes of point defects in solids is a
topic with long
history~\cite{Huang1950,markham1959,Osadko1979,Stoneham}. Yet,
atomistic first-principles calculations~\cite{Martin} have been
difficult. Certain properties of defects, such as discrete energy
levels or localized magnetic moments, resemble those of atoms or
molecules. However, those localized states are surrounded by a huge
number of other electrons and ions, which makes point defects
qualitatively different from atoms and molecules and requires a
treatment as a solid-state system. Unfortunately, unlike perfect
crystals, defects do not possess translational symmetry, which
significantly complicates their quantum-mechanical \mbox{description}.

Optical signatures of defects in semiconductors and insulators are
properties where both the atomic and the solid-state aspects are
manifest. Many defects have localized levels in the band gap of the
host material, on par with atoms or molecules in vacuum. Optical
transitions involving those levels usually lead to lattice
rearrangement. This rearrangement typically couples to a continuum of
vibrational modes with different frequencies, in contrast to
molecules, where a finite number of vibrations participate. As a
result, optical signatures of defects are composed of continuous
\textit{bands}~\cite{Stoneham}.

Due to the complexity of the problem earlier first-principles calculations
of optical bands involved approximations that were not fully tested.
Kretov~\textit{et~al.}~\cite{Kretov2012} performed a study of the
luminescence lineshape of a Mn impurity in
$\mathrm{Zn}_2\mathrm{SiO}_4$, but approximated the vibrational modes
of the defect system by those of the bulk crystal.
Alkauskas~\textit{et al.} calculated luminescence lineshapes of
defects in GaN and ZnO with a very strong electron--phonon coupling
(as~quantified by the so-called Huang--Rhys factor~\cite{Huang1950}
$S\gg1$, discussed below)~\cite{alkauskas2012first}. For such systems
optical lineshapes can be accurately calculated by mapping all
vibrations onto an effective one-dimensional vibrational
problem~\cite{Huang1950}. In subsequent work, Alkauskas
\textit{et~al.} calculated the luminescence lineshape for a negatively
charged nitrogen--vacancy (NV) center in diamond~\cite{Alkauskas2014},
a defect in which electron--phonon coupling is not strong enough to
justify the single-mode approximation. They explicitly treated all
phonons pertaining to the defect system by diagonalizing dynamical
matrices of periodically repeated supercells~\cite{Freysoldt2014} with
up to 11\,000 atoms. The calculations took into account only symmetric
$a_1$ vibrational modes, whereas it is known that asymmetric $e$ modes
also contribute~\cite{Doherty2013} due to the 
dynamical Jahn--Teller (JT) effect in the electronically excited
state~\cite{Fu2009}. A few other recent studies have calculated
optical lineshapes of point
defects~\cite{thiering2017ab,Hashemi2020,Bouquiaux2020}. Most
of the published work used fairly small supercells, which yields an
incomplete description of the phonon spectrum. It~can be concluded
that the status of calculations of optical lineshapes of point defects
defects is much behind that of molecules where very accurate
calculations are now being routinely
performed~\cite{Berger1998,Dierksen2004,Borrelli2013}.

A proper inclusion of the dynamical JT effect is very difficult.
Historically, an understanding of this effect has been achieved by
invoking an effective single-mode approximation for the degenerate
asymmetric vibrational mode~\cite{obrien1972}. This
approximation~suffices to obtain the general features of optical
lineshapes~\cite{Davies1981} or to assess the effect of the JT
interaction on spin--orbit coupling~\cite{Ham1968}. However, 
the single-mode approach is insufficient in the case of defects
with weak to moderate electron--phonon coupling. Unfortunately, the
problem of diagonalizing the vibronic (\textit{i.e.}, combined
vibrational and electronic) Hamiltonian has an extremely unfavorable
scaling with regard to the number of vibrational modes that are
included. In the
past, this limited the consideration of the dynamical JT effect to no
more than just a few vibrational degrees of freedom~\cite{obrien1980}.

In this work, we present first-principles calculations of defect
optical lineshapes for the NV center in diamond. The NV center is a
very useful model system because its structure is accurately known and
the defect is extremely well characterized spectroscopically. Our
present work goes well beyond the ideas developed in
Ref.~\onlinecite{Alkauskas2014} and significantly advances the
methodology and calculations in the following areas: (i)~We present
the calculations of absorption lineshapes in addition to luminescence
lineshapes. (ii)~The contribution of asymmetric $e$ vibrations is
included. The latter is possible since (iii)~we propose and test an
efficient algorithm to solve the multi-mode Jahn--Teller problem, which
allows us to treat vibronic coupling to a very large number of
asymmetric modes.

Our work is not just an academic exercise in the development of
quantum-mechanical methods for solid-state systems. Recently, the
field of point defects in solids has experienced a renaissance due to
the application of point defects in various branches of quantum
technologies: quantum computing, quantum communication, and quantum
sensing~\cite{awschalom2018}. While the NV center is probably one of
the most prominent examples of these so-called quantum defects, many
other defects have been addressed~\cite{awschalom2018}. There is a
wealth of experimental data on various point defects in solids and
many potentially useful systems await to be discovered or identified.
The ability to accurately calculate optical signatures of point
defects will greatly aid their identification and potential
application. We note that the dynamic Jahn--Teller effect influences not
only the vibrational sidebands but also the fine structure of the  
spectra, in particular the splitting of the zero-phonon line (ZPL, see below
for a definition). In the case of the NV center the fine structure is affected
by the spin-orbit coupling in the excited state, which, in its turn, is reduced
when the dynamic Jahn-Teller effect is present \cite{Ham1968}. The
fine structure of the spectra will not be considered in this work.

The paper is organized as follows. In Sec.~\ref{sec:nv} we introduce
the nitrogen--vacancy center. In Sec.~\ref{sec:th} the general theory
of optical lineshapes of defects is laid down. The first-principles
methodology and actual calculations are presented in
Sec.~\ref{sec:first-princ}. These calculations are subsequently used
to build effective models of NV centers in very large supercells, as
detailed in Sec.~\ref{sec:embedd}. Coupling to fully symmetric $a_1$
vibrational modes is discussed in Sec.~\ref{a1} and the interaction
with $e$ modes, \textit{i.e.}, the dynamical multi-mode Jahn--Teller
effect, is analyzed in Sec.~\ref{jt}. The calculated lineshapes, where
the contributions of $a_1$ and $e$ modes are combined, are presented
and compared to experiments in Sec.~\ref{sec:comb}. Our findings are
summarized in Sec.~\ref{sec:disc}, where we also discuss possible
sources of the remaining small discrepancies between theory and
experiment.


\begin{figure*}
\includegraphics[width=0.9\textwidth]{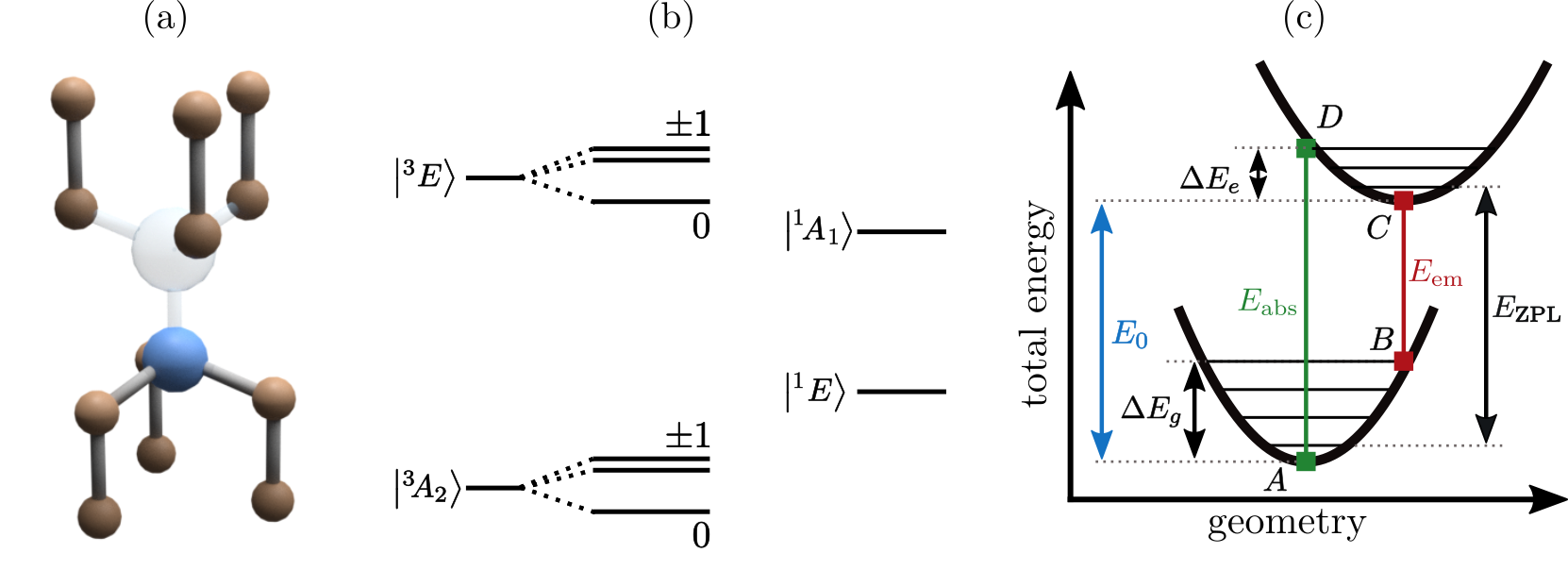}
\caption{(a)~Atomic structure of the NV center in diamond.
  (b)~Energy-level diagram of the NV center. Splitting of the magnetic
  sub-levels is not to scale. (c)~One-dimensional representation of
  adiabatic potential energy surfaces in the ground and the excited
  states.  Points $A$ and $D$ represent equilibrium geometry of the
  ground state, while points $B$ and $C$ represent equilibrium
  geometry of the excited state. $\Delta E_g$ and $\Delta E_e$ are
  lattice relaxation energies. $E_0$ is energy difference between the
  potential energy minima, and $E_{\text{ZPL}}$ is the energy of the
  zero-phonon line.\label{fig:nv}}
\end{figure*}

\section{The nitrogen--vacancy center\label{sec:nv}}

The defect studied in this paper is the negatively charged
nitrogen--vacancy (NV) center in diamond, depicted in
Fig.~\ref{fig:nv}(a). It is a complex of a substitutional nitrogen
atom with a carbon vacancy; the extra electron comes from remote
donors. Over the past two decades the NV center has been the focus of
a lot of research activity~\cite{Doherty2013} due to its application
in nanoscale sensing~\cite{Schirhagl2014}, quantum
communication~\cite{Childress2013}, and quantum
computation~\cite{Waldherr2014}.

The point-group symmetry of the defect is $C_{3v}$ with a three-fold
axis going through the nitrogen and the vacant site. The energy level
diagram of the negatively charged NV center is shown in
Fig.~\ref{fig:nv}(b). The NV center has a triplet ground state $\tA2$
and a triplet excited state~$\tE$. Once the system is in the excited
state, it can either return to the ground state via a radiative
transition or undergo an inter-system crossing to the singlet state
$\sA1$. The latter process is important for optical spin polarization
and read-out; for more details see Ref.~\onlinecite{Doherty2013}. In
this paper, we will focus on optical transitions between $\tE$ and
$\tA2$ states. Also, we will use the term ``NV center'' omitting the
explicit mention of its negative charge state.

The role of lattice vibrations in optical properties can be
qualitatively illustrated using an effective one-dimensional (1D)
representation, the so-called configuration coordinate
diagram~\cite{Stoneham,Alkauskas2016tutorial}, shown in
Fig.~\ref{fig:nv}(c). The equilibrium geometries of the defect in the
excited and the ground state are different, which is reflected by the
fact that total energies in different electronic states attain their
minimum for different configurations. In the so-called classical
Franck--Condon picture the energy $E_{\text{abs}}$ is associated with
absorption energy and the energy $E_\text{em}$ with emission. Lattice
relaxation energies during optical transitions are quantified by
Franck--Condon shifts in the ground ($\Delta E_g$) and the excited
($\Delta E_e$) state.

Let the energy difference between the equilibrium configurations be
$E_0$ and let $\varepsilon_{gn}$ and $\varepsilon_{em}$ be energy
eigenvalues of quantum-mechanical vibronic levels in the ground and
the excited state, referenced to the respective potential energy
minima. The optical transition between the lowest vibronic level in
the excited state and the lowest vibronic level in the ground state is
called the zero-phonon line (ZPL); its energy is
\begin{equation}
E_{\text{ZPL}}=E_0 + \varepsilon_{e0} - \varepsilon_{g0}\,.
\label{eq:ZPL}
\end{equation}
The zero-phonon line is the transition during which the net number of
phonons in the system does not change. Strictly speaking, the
definition Eq.~\eqref{eq:ZPL} is valid only for $T=0$~K. In this paper
we will assume this low-temperature limit and calculate luminescence
and absorption lineshapes for $T=0$~K.

Because of different geometries in the excited and the ground state,
luminescence and absorption signals are not narrow lines at the
energy of the ZPL, but become broadened due to lattice vibrations,
resulting in bands. A~quantitative theory of the vibrational
structure of optical lines is given next, in Sec.~\ref{sec:th}.


\section{Vibrational broadening of optical transitions\label{sec:th}}

The process of optical emission is described by an energy-dependent
emission probability $P(\hbar\omega)$, number of photons emitted in
the solid angle $4\pi$ per unit energy per unit time. The process of
absorption is described by an absorption cross section
$\sigma(\hbar\omega)$. These quantities are given by~\cite{Lax1952}:
\begin{eqnarray}
  P(\hbar\omega)=C_{P}\omega^{3}I_{\text{em}}(\hbar\omega);
  \quad
  \sigma(\hbar\omega)=C_{\sigma}\omega I_{\text{abs}}(\hbar\omega),
\label{eq:general1}
\end{eqnarray}
where
\begin{align}
  \label{eq:general_spectrum1}
  & \hspace{-2pt}
    I_{\text{em}}(\hbar\omega)
    \!=\!
    \sum_{n}
    \left|\left\langle
    \Psi_{e0}\left|\dipole\right|\Psi_{gn}
    \right\rangle \right|^{2}
    \!\delta\!
    \left(
    E_0+\varepsilon_{e0}-\varepsilon_{gn}-\hbar\omega
    \right)\!,
\end{align}
and
\begin{align}
  \label{eq:general_spectrum2}
  & \hspace{-2pt}
  I_{\text{abs}}(\hbar\omega)
  \!=\!
  \sum_{n}
  \left|\left\langle
  \Psi_{g0}\left|\dipole\right|\Psi_{en}
  \right\rangle\right|^{2}
  \!\delta\!
    \left(E_0+\varepsilon_{en}-\varepsilon_{g0}-\hbar\omega\right)
    \!.
\end{align}
In the subscripts $g$ stands for ``ground'' and $e$ for ``excited''.
$C_{P}$ and $C_{\sigma}$ contain fundamental physical constants, as
well as the refractive index of diamond, $n_r$. As we will be
interested in normalized lineshapes, these constants will not be
discussed further.  Wavefunctions $\left| \Psi_{gn} \right\rangle$ and
$\left|\Psi_{en} \right\rangle$ describe vibronic states in the ground
and excited electronic manifolds; their energies $\varepsilon_{gn}$
and $\varepsilon_{en}$ were introduced in Sec.~\ref{sec:nv}.
$\dipole$ is the dipole operator; in the most general case it can be
written as the sum of electronic and ionic contributions:
\begin{eqnarray}
  \dipole = \dipole_{el} + \dipole_N.
\label{dipole}
\end{eqnarray}

In this paper we will employ the so-called crude adiabatic (also
called static) approximation. In this approximation electronic
wavefunctions depend only on electronic degrees of freedom and they
are eigenfunctions of the electronic Hamiltonian corresponding to some
fixed position of ions~\cite{azumi1977does}.

The ground state of the NV center is an orbital
singlet~\cite{Doherty2013}, and its vibronic states can be written as:
\begin{equation}
  \gwf[][g;pr] = \vwf[g][p][a_1][a_1] \vwf[g][r][e][e] \ket{A_2}.
\label{WF1}
\end{equation}
Here the $a_1$ and $e$ symmetry components of the vibrational
wavefunction are shown explicitly. To describe these components we
need two quantum numbers $p$ and $r$ that in the previous expressions
were substituted with a single index $n$.  To avoid possible confusion
the index `$e$' for `excited' will be used as a subscript, and the
index `$e$' to label the $e$ irreducible representation will be used
as a superscript whenever both indices appear on the same symbol.

The excited state is an orbital doublet~\cite{Doherty2013}, and the
general expression of its vibronic states is~\cite{bersuker2012}:
\begin{equation}
  \gwf[][e;st] = \vwf[e][s][a_1][a_1]
  \left[
    \vwf[e][t][e_x][e]\ket{E_x} +
    \vwf[e][t][e_y][e]\ket{E_y}
  \right],
  \label{WF2a}
\end{equation}
where $s$ and $t$ are used to label ionic wavefunctions of $a_1$ and
$e$ symmetry, respectively; like for the ground state, quantum numbers
$s$ and $t$ replace a single quantum number $n$ in
Eq.~\eqref{eq:general_spectrum2}.  Using the notation
\begin{equation}
  \ket{\Phi_{et}} =
  \vwf[e][t][e_x][e]\ket{E_x} +
  \vwf[e][t][e_y][e]\ket{E_y}
  \label{eq:vibronic_term}
\end{equation}
we can rewrite the vibronic wavefunction in the excited state as:
\begin{equation}
  \gwf[][e;st] = \vwf[e][s][a_1][a_1]\ket{\Phi_{et}}.
  \label{WF2}
\end{equation}
This form will be useful in Sec.~\ref{jt}.

In the expressions above ionic wavefunctions pertaining to vibrational
modes of $a_2$ symmetry are not explicitly shown. In the case of the
NV center these modes show up neither in absorption nor emission. The
reason for this is that the difference between equilibrium geometries
in the $\tA2$ and $\tE$ electronic states (to be discussed in more
detail in Secs.~\ref{a1} and~\ref{jt}) contains only $a_1$ and $e$
components, but no $a_2$ components. I.e., the projection of that
difference on $a_2$ vibrational modes is zero. Thus, these modes will
not be considered further in this work. However, $a_2$ modes can in principle
contribute in some other $C_{3v}$ systems. As these modes do not break
the degeneracy of $E$ electronic states, they are not Jahn-Teller active.
Therefore, they can be treated the same way as $a_1$ modes. 

When ground and excited state wavefunctions are in the form of
Eqs.~\eqref{WF1} and~\eqref{WF2a}, the nuclear part of the dipole
operator [Eq.~\eqref{dipole}] does not contribute.  If the symmetry
axis of then NV center is along the $z$ direction, then light emitted
from this center is polarized in the $xy$ plane.  Using
group-theoretical analysis one can then show that
\begin{eqnarray}
  && \left|
  \matrixel{\gwf*[][g;pr]}{\dipole_{el}}{\gwf*[][e;st]}
  \right|^2 \\
    \label{eq:matrix_el}
  &&\quad = |\mu_0|^2
     \left|\braket{\vwf*[g][p][a_1][a_1]}{\vwf*[e][s][a_1][a_1]}\right|^2
     \left[
     \left|\braket{\vwf*[g][r][e][e]}{\vwf*[e][t][e_x][e]}\right|^2 +
     \left|\braket{\vwf*[g][r][e][e]}{\vwf*[e][t][e_y][e]}\right|^2
     \right]
     \nonumber
\end{eqnarray}
where $\mu_0=\sum_{i}e\matrixel{A_2}{x_i}{E_y}$ is the reduced matrix
element in the Wigner--Eckart theorem (the sum runs over all electrons
of the negatively charged NV center and $e$ is elementary charge).
Physically, $\mu_0$ is simply the transition dipole moment.

The energies of vibronic levels that appear in
Eqs.~\eqref{eq:general_spectrum1} and~\eqref{eq:general_spectrum2} are
a sum of contributions by $a_1$ and $e$ modes, {\it i.e.}
$\varepsilon_{gn}=\varepsilon^{a_1}_{gp}+\varepsilon^{e}_{gr}$ and
similarly for the excited state. In this case one can show that
functions $I_{\text{em}}$ and $I_{\text{abs}}$ from
Eqs.~\eqref{eq:general_spectrum1} and~\eqref{eq:general_spectrum2} can
be expressed as:
\begin{eqnarray}
  I_{\{\text{em,abs}\}}(\hbar\omega) =
  |\mu_0|^2 A_{\{\text{em,abs}\}}(\hbar\omega),
\label{I}
\end{eqnarray}
where $A(\hbar\omega)$, the spectral function, is given by the
following expression:
\begin{eqnarray}
  A_{\{\text{em,abs}\}}(\hbar\omega)
  =
  \!\int\!\!
  A_{a_{1}}\!(\hbar\omega-\hbar\omega')
  A_{e}(\hbar\omega')\,
  \mathrm{d}(\hbar\omega').
\label{conv}
\end{eqnarray}
This is a convolution of the two spectral functions pertaining to
$a_1$ and $e$ modes.  In the case of luminescence we
define these spectral functions by:
\begin{equation}
  A_{a_1}(\hbar\omega) =
    \sum_p
    \left|
    \braket{\vwf*[e][0][a_1][a_1]}{\vwf*[g][p][a_1][a_1]}
    \right|^2
    \delta
    \left(
     E_{\mathrm{ZPL}} + \varepsilon^{a_1}_{g0} - \varepsilon^{a_1}_{gp} -\hbar\omega
    \right)
\label{a1-lum}
\end{equation}
and
\begin{align}
  A_e(\hbar\omega) =
  & \sum_r
    \left[
    \left|\braket{\vwf*[e][0][e_x][e]}{\vwf*[g][r][e][e]}\right|^2 +
    \left|\braket{\vwf*[e][0][e_y][e]}{\vwf*[g][r][e][e]}\right|^2
    \right]
    \notag \\
  & \times
    \delta(\varepsilon^{e}_{g0} - \varepsilon^e_{gr} - \hbar\omega) \, .
\label{e-lum}
\end{align}
In the case of absorption they are:
\begin{equation}
  A_{a_1}(\hbar\omega) =
  \sum_s
  \left|
    \braket{\vwf*[e][s][a_1][a_1]}{\vwf*[g][0][a_1][a_1]}
  \right|^2\delta
  \left(
    E_{\mathrm{ZPL}} + \varepsilon^{a_1}_{es} - \varepsilon^{a_1}_{e0} - \hbar\omega
  \right)
\label{a1-abs}
\end{equation}
and
\begin{eqnarray}
  A_e(\hbar\omega) =
  &
    \sum_t
    \left[
    \left|\braket{\vwf*[e][t][e_x][e]}{\vwf*[g][0][e][e]}\right|^2 +
    \left|\braket{\vwf*[e][t][e_y][e]}{\vwf*[g][0][e][e]}\right|^2
    \right]
    \notag
  \\
  & \times \delta(\varepsilon^{e}_{et} - \varepsilon^{e}_{e0} - \hbar\omega) \, .
\label{e-abs}
\end{eqnarray}
Since $A_{a_1}$ carries the lion's share of luminescence and
absorption lineshapes (as discussed in Secs.~\ref{a1},~\ref{jt},
and~\ref{sec:comb}), we choose to define $A_{a_1}$ with respect to
$E_{{\rm ZPL}}$, while $A_e$ is defined with respect to energy
$0$. This is done for convenience only and other choices are certainly
possible. Therefore, in the case of emission $A_{a_1}(\hbar\omega)$ is
non-zero for energies smaller than $E_{\text{ZPL}}$, while
$A_{e}(\hbar\omega)$ is non-zero for energies smaller than 0. In the
case of absorption $A_{a_1}(\hbar\omega)$ is nonzero for energies
larger than $E_{\text{ZPL}}$, and $A_{e}(\hbar \omega)$ is nonzero for
energies larger than 0. All quantities $A(\hbar \omega)$ are
automatically normalized to $1$ as per their definition.

As mentioned above, in this work we will not deal with absolute
luminescence intensities and absorption cross-sections, defined in
Eq.~\eqref{eq:general1}, but rather with normalized
lineshapes. Comparing with Eq.~\eqref{eq:general1}, we see that those
are given by:
\begin{equation}
  L_{\text{em}}(\hbar\omega)=N_1\omega^3A_{\text{em}}(\hbar\omega)
  \label{Lem}
\end{equation}
for emission and
\begin{equation}
  L_{\text{abs}}(\hbar\omega)=N_2\omega A_{\text{abs}}(\hbar\omega)
\label{Lab}
\end{equation}
for absorption, where $N_1$ and $N_2$ are normalization constants,
needed because of the appearance of factors $\omega^3$ and $\omega$ in
$L_{\{\text{em,abs}\}}(\hbar\omega)$. The principal task of the
current paper is the evaluation of Eqs.~\eqref{conv}--\eqref{Lab} for
the NV center.


\section{First-principles calculations\label{sec:first-princ}}

Calculations have been performed within the framework of density
functional theory (DFT). Exchange and correlation was described by the
hybrid functional of Heyd, Scuseria, and Ernzerhof
(HSE)~\cite{Heyd2003}. In HSE a fraction $a=1/4$ of screened Fock
exchange is admixed to the semi-local exchange based on the
generalized gradient approximation (GGA) in the form of Perdew, Burke,
and Ernzerhof (PBE)~\cite{Perdew1996}. As discussed below, the HSE
functional provides a very good description of the electronic
structure of bulk diamond (in particular its band~gap), and optical
excitations energy of the NV center~\cite{Gali2009}
(Sec.~\ref{subsec:nv}). However, the PBE functional yields better
agreement with experiment for the bulk lattice constant, the bulk
modulus, and the vibrational properties of diamond~\cite{Hummer2009}
(Sec.~\ref{subsec:bulk_diamond}). For this reason we also perform
calculations at the PBE level. We used the projector-augmented wave
approach with a plane-wave energy cutoff of 500 eV. Other
computational details are given when describing specific
systems. Calculations have been performed with the Vienna Ab-initio
Simulation Package ({\sc vasp})~\cite{VASP}.

\subsection{Bulk parameters and lattice vibrations\label{subsec:bulk_diamond}}

Lattice relaxation was performed using a conventional cubic cell with
eight carbon atoms.  The Brillouin zone was sampled using the
Monkhorst--Pack~\cite{Martin} $8\times8\times8$ $k$-point
mesh. Results are summarized in Table~\ref{tab:lattice}.  We observe
that PBE provides a better description of the lattice constant and the
bulk modulus.  However, the band gap is significantly underestimated
in PBE ($E_g=4.12$~eV, compared to the experimental value of
$5.48$~eV~\cite{Madelung2012}). The band gap is much closer to
experiment in HSE ($E_g=5.36$~eV).

\newcounter{auxFootnote}
\begin{table}
  \caption{\label{tab:lattice}Calculated lattice constants $a$ (\AA),
    bulk moduli $B$ (GPa), and highest phonon frequencies (in meV) at
    high-symmetry points in diamond. Experimental
    values~\cite{Madelung2012,Zouboulis1998,Warren1967} are listed for
    comparison. For the lattice constant and bulk modulus the
    deviation from the experimental value is indicated in
    parentheses.}
  \begin{ruledtabular}
    \begin{tabular}{l@{\hspace{-5.2em}}ddddd}
      & \multicolumn{1}{c}{$a$}
      & \multicolumn{1}{c}{$B$}
      & \multicolumn{1}{c}{$\omega(\Gamma)$}
      & \multicolumn{1}{c}{$\omega(\mathrm{X})$}
      & \multicolumn{1}{c}{$\omega(\mathrm{L})$}
      \vspace{2pt}
      \\
      \hline\\[-1.5ex]
      PBE & 3.574\ (+0.20\%)  & 430\ (-0.19\%) & 160.5 & 147.6 & 154.2 \\
      HSE & 3.548\ (-0.53\%)  & 470\ (+0.61\%) & 169.9 & 155.1 & 161.1 \\
      expt.
      & 3.567\footnotemark[1]
      & 443\footnotemark[2]
      & 166.7\footnotemark[3]
      & 149.2^{\text{b}}
      & 153.0^{\text{b}}
      \\
    \end{tabular}
  \end{ruledtabular}
\footnotetext[1]{Reference \onlinecite{Madelung2012}}
\footnotetext[2]{Reference \onlinecite{Zouboulis1998}}
\footnotetext[3]{Reference \onlinecite{Warren1967}}
\end{table}


\begin{figure}
\includegraphics[width=1\linewidth]{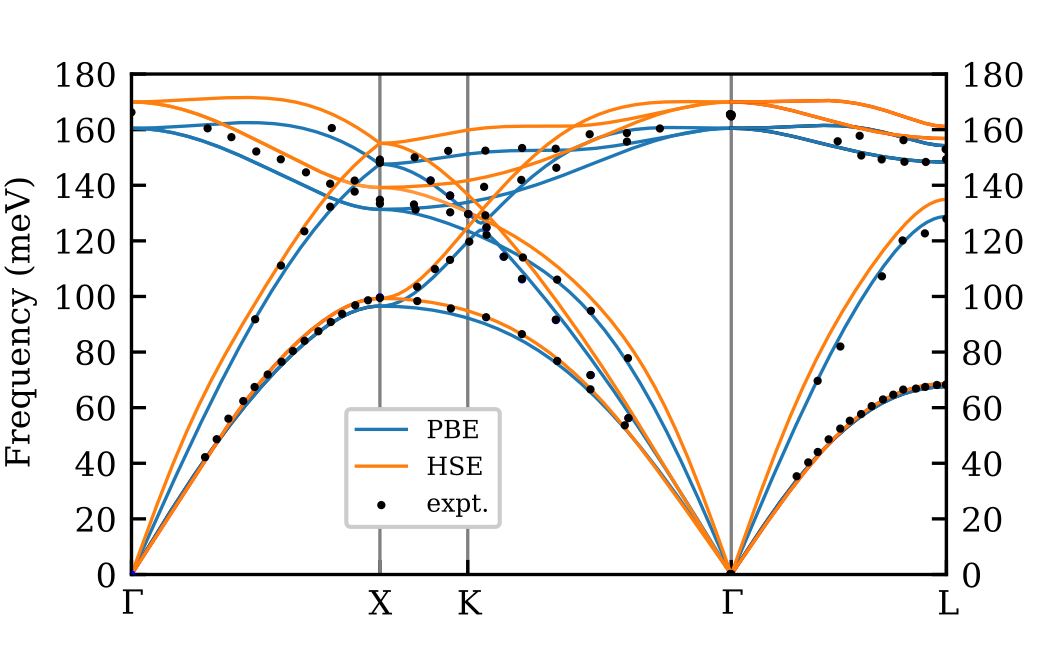}
\caption{Phonon dispersion curves of diamond calculated using PBE
  (blue lines) and HSE (red lines) functionals. Experimental values
  are taken from Ref.~\onlinecite{Warren1967}.\label{pdispersion}}
\end{figure}

Phonon dispersion curves were calculated using the \texttt{PHONOPY}
software package~\cite{phonopy}. Force constants were computed in
$4 \times 4 \times 4$ supercells (containing 512 atoms) using finite
displacements and a single $k$-point (at $\Gamma$) for the Brillouin-zone
sampling.  We have used displacements $\Delta = 0.01$ \AA.\@ In
Fig.~\ref{pdispersion} the calculated phonon dispersion curves are
compared with inelastic neutral scattering data from
Ref. \onlinecite{Warren1967}.  Our calculations represent phonon
dispersion for $T=0$, while experiments have been performed at room
temperature. However, due to the very rigid nature of the diamond
lattice the phonon frequencies in diamond change by less than 0.1 meV
from cryogenic temperatures to room temperature~\cite{solin1970},
justifying the comparison of calculations with room-temperature data.
Our calculations are in full agreement with those of
Ref.~\onlinecite{Hummer2009}. Both PBE and HSE functionals describe
the phonon dispersion reasonably well. However, PBE provides a
slightly more accurate description of the spectrum, \textit{e.g.}, for
longitudinal acoustic phonons or optical phonons along the paths
$\rm{XK}$, $\rm{K\Gamma}$, $\rm{\Gamma L}$ (Fig.~\ref{pdispersion}).
The highest phonon frequencies for high-symmetry points are compared
with experiment in Table~\ref{tab:lattice}.

\subsection{Negatively charged NV center\label{subsec:nv}}

In this Section we discuss calculations of the NV center and analyze
the convergence of these calculations with respect to the supercell
size. We also present the calculations of the vibrational properties
of the NV center.

\subsubsection{Ground state}

First-principles calculations of the ground and the excited state of
the NV center have been reported previously (for a review see
Ref.~\onlinecite{Gali2019}). Here we give an of overview of our
calculations for completeness. The NV center possesses a $C_{3v}$
point symmetry and has a stable paramagnetic ground state $\tA2$ with
spin $S=1$.

\begin{figure}
\includegraphics[width=8.5cm]{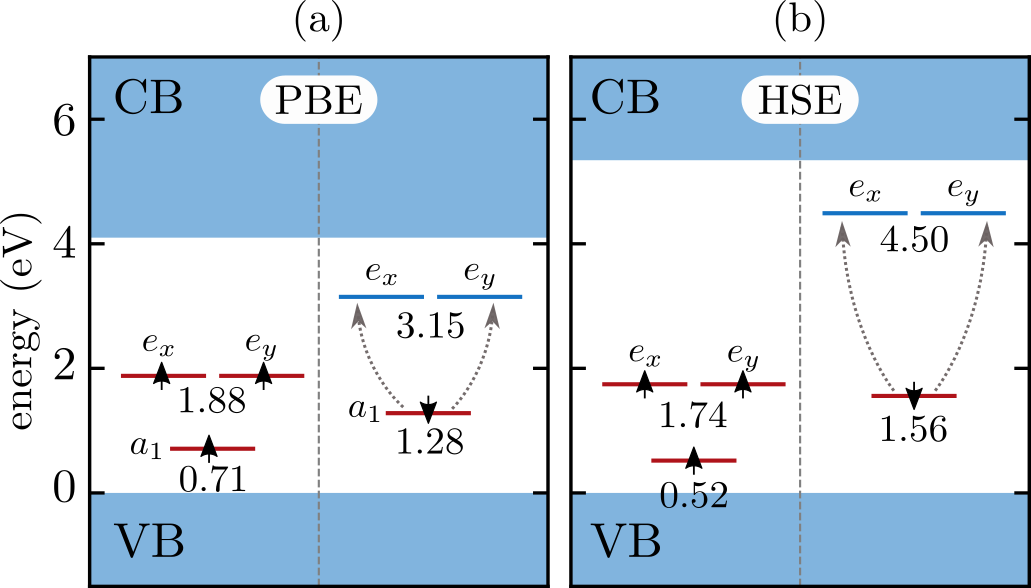}
\caption{Defect-level diagrams of the NV$^{-1}$ center calculated
  using PBE (a) and HSE (b) functionals. Diagrams show Kohn--Sham
  single particle defect levels for the ground state $\tA2$. The
  spin-majority channel is denoted with upward arrows and the
  spin-minority channel with downward arrows. Shaded areas correspond
  to the valence band (VB) and the conduction band (CB).
  Dotted arrows show optical excitation.
  \label{fig:defect-levels}}
\end{figure}

Calculated Kohn--Sham defect level diagrams are shown in
Fig.~\ref{fig:defect-levels} for both functionals.  In these
calculations the defect supercell is a $4\times4\times4$ cell with 512
atomic sites.  We used a single $k$-point (at $\Gamma$) for
Brillouin-zone sampling. Theoretical lattice constants were used
(Table~\ref{tab:lattice}) consistently for both functionals. For both
spin channels there are three defect levels in the band gap: a~fully
symmetric $a_1$ level and a doubly-degenerate $e$ level. All the
levels are filled in the spin-majority channel, and only the $a_1$
level is filled in the spin-minority channel~\cite{Doherty2013}.  The
splitting between the two sets of levels is larger in HSE:\@ in the
spin-minority channel the difference between the $a_1$ level and the
$e$ levels is 1.87~eV for PBE and 2.94~eV for HSE.\@ The $\tA2$ ground
state can be described by the electronic configuration $a_1^2e^2$.

\subsubsection{Excited state {\label{sec:exc}}}

The excited state triplet $\tE$ is obtained by promoting the electron
in the spin-minority channel from the $a_1$ level to the $e$ level (as
shown by dashed arrows in Fig.~\ref{fig:defect-levels}), resulting in
the electronic configuration $a_1e^3$.  We treat the electronic
structure of the excited state using the $\Delta$SCF
method~\cite{Jones1989}, whereby the state is modeled by constraining
appropriate Kohn--Sham orbital occupations.  The $\Delta$SCF method is
originally due to Slater, who applied it to optical excitation in
atoms; the method was first applied to the NV center by Gali {\it et
  al.}~\cite{Gali2009}

The $\tE$ state is a $E \otimes e$ Jahn--Teller (JT)
system~\cite{Fu2009,abtew2011dynamic} (more precisely, it is a
$E \otimes (e \oplus e \cdots)$ system). The reason for the
Jahn--Teller instability is the degeneracy of nominal $a_1e_x^2e_y^1$
and $a_1e_x^1e_y^2$ configurations of the $\tE$ manifold. The $\tE$
state is unstable with respect to symmetry breaking due to the
interaction with $e$ phonons, leading to the lowering of the
energy. The Jahn--Teller effect is
\textit{dynamical}~\cite{Fu2009,abtew2011dynamic}, which qualitatively
means that on average the NV center still retains $C_{3v}$
symmetry. I.e., the adiabatic potential energy surface has a minimum
off the high-symmetry point, but the total wavefunction of the ground
vibronic level is totally symmetric.  The fundamental consequence of
the dynamical nature is that wavefunctions that describe the entire
system of ions and electrons cannot be factorized into the ionic part
and the electronic part. This is already expressed by
Eqs.~\eqref{WF2a} and~\eqref{eq:vibronic_term}. A more rigorous
discussion of the dynamical vs.~the static JT effect can be found in
Ref.~\onlinecite{bersuker2012}.

Vibrational modes obtained in the JT-distorted geometry are
inconvenient for theoretical analysis because they lack symmetry
properties of the $C_{3v}$ point group.  The existing understanding of
the Jahn--Teller effect relies on the use of vibrational modes that
have the symmetry of the system in the undistorted configuration,
$C_{3v}$ in our case~\cite{bersuker2012}.  Unfortunately, calculations
of vibrational frequencies in $C_{3v}$ symmetry for degenerate states
with configurations $a_1e_x^2e_y^1$ or $a_1e_x^1e_y^2$ pose
computational difficulties due to the degeneracy of these states.  A
correct way to proceed would be to perform vibrational analysis for a
mixed electronic state, the density of which is the average of charge
densities of the $a_1e_x^2e_y^1$ and $a_1e_x^1e_y^2$ configurations.
One can expect that the charge density of such a mixed state would be
well approximated with a system with a configuration
$a_1e_x^{1.5}e_y^{1.5}$, whereby the electron in the spin-minority
channel is split between $e_x$ and $e_y$ levels. It is this state with
fractional occupations that is used here for the calculation of
vibrational properties in the excited state.  From now on, the
equilibrium geometry of the excited state with the electronic
occupations $a_1e_x^{1.5}e_y^{1.5}$ will be called a symmetric
excited-state configuration (it retains the $C_{3v}$ symmetry). The
equilibrium geometry with electronic occupations $a_1e_x^2e_y^1$ or
$a_1e_x^1e_y^2$ will be called an asymmetric excited-state
configuration. The asymmetric configuration corresponds to a true
physical minimum on the adiabatic excited-state potential energy
surface.

\subsubsection{Optical excitation energies {\label{sec:exc_en}}}

We first test the convergence of calculated energy differences between
the ground and excited states $E_0$, as well as lattice relaxation
energies $\Delta E_e$ and $\Delta E_g$ [Fig.~\ref{fig:nv}(c)] as a
function of the supercell size. In these calculations the excited
state wavefunction and geometry corresponds to the symmetric electron
configuration {\sym}. We note that the energy of the system with this
electron configuration is not strictly physical due to Coulomb
repulsion between a ``split'' electron occupying two different
levels. To emphasize this fact we add a ``prime'' to energies
calculated with this occupation ($E_0'$,~$\Delta
E_e'$,~$\Delta E_g'$). These calculations serve as a numerical test
regarding convergence of calculated energies as a function of the
supercell size, as they are faster than the calculation with the
actual asymmetric configuration $a_1e_x^2e_y^1$ or
$a_1e_x^1e_y^2$. For the same reason we employ the PBE functional in
these tests. Results for supercells $3\times 3\times 3$ (216 atomic
sites), $4\times 4\times 4$ (512 sites), and $5\times 5\times 5$ (1000
sites) are shown in Table~\ref{tab:conv}. The Brillouin zone was
sampled at the $\Gamma$ point. From the results we see that the
$4\times 4\times 4$ supercell is sufficient to describe energies with
an accuracy of 0.01 eV.

\begin{table}[t]
  \caption{Convergence of the energy difference between the
    equilibrium configurations $E_0'$, as well as lattice relaxation
    energies $\Delta E_g'$ and $\Delta E_e'$ as a function of the
    supercell size (in eV).  Values correspond to the PBE functional
    and the symmetric electron configuration {\sym} in the excited
    state.\label{tab:conv}}
  \begin{ruledtabular}
    \begin{tabular}{lccc}
      & $3 \times 3 \times 3$ & $4 \times 4 \times 4$ & $5 \times 5 \times 5$ \\
      \hline
      \\[-2ex]
      $E_{0}'$       & 1.719 & 1.694 & 1.689  \\
      $\Delta E_g'$  & 0.155 & 0.173 & 0.174  \\
      $\Delta E_e'$  & 0.185 & 0.199 & 0.205
    \end{tabular}
  \end{ruledtabular}
\end{table}

\begin{table}[t]
  \caption{Calculated $E_0$ and relaxation energies for the PBE and
    HSE functionals (in eV). Values correspond to the electronic
    configuration {\asym} in the excited state and the
    $4\times 4 \times 4$ supercell. In parentheses $a_1$ and $e$
    components of Franck--Condon shifts are given.
    }\label{tab:conv2}
  \begin{ruledtabular}
    \begin{tabular}{lcc}
                                   & PBE                    & HSE \\
      \hline
      \\[-1.5ex]
      $E_0$                        & 1.689                  & 1.995  \\
      $\Delta E_g$ ($a_1$ + $e$)   & 0.196 (0.159 + 0.037)  & 0.257 (0.214 + 0.043) \\
      $\Delta E_e$ ($a_1$ + $e$)   & 0.218 (0.182 + 0.036)  & 0.298 (0.256 + 0.042)  \\
    \end{tabular}
  \end{ruledtabular}
\end{table}

In Table~\ref{tab:conv2} the calculations of $E_0$, $\Delta E_g$, and
$\Delta E_e$ with PBE and HSE functionals are summarized. Values
correspond to the physical (asymmetric) electron configuration {\asym}
in the excited state and pertain to the $4 \times 4 \times 4$
supercell.  Franck--Condon shifts $\Delta E_g$ and $\Delta E_e$ occur
due to lattice relaxation which, as discussed in Sec.~\ref{sec:th} has
$a_1$ and $e$ components. These two contributions are also presented.
In short,  in the case of $a_1$ modes this is given by $\sum_k S_k \hbar \omega_k$;
in the case of $e$ modes this is given by $\sum_k K_k^2 \hbar \omega_k$
(see Secs.~\ref{a1} and \ref{jt} for the definition of the quantities $S_k$ and $K_k^2$). 
We note that the $e$ component of $\Delta E_e$ is exactly the
Jahn--Teller relaxation energy, often labelled $E_{\text{JT}}$.  The
value of $E_0$ calculated in HSE, 1.995 eV, is close to the
experimental ZPL of 1.945 eV. Note that in order to determine the
theoretical value of the ZPL we would need to add the contributions of
zero-point vibrations $\varepsilon_{e0}-\varepsilon_{g0}$ as per
Eq.~\eqref{eq:ZPL}. Typically this contribution is of the order of
$\sim$10 meV~\cite{Londero2018}. We can conclude that the HSE
functional provides a very good description of optical excitation
energies.

\subsubsection{Vibrational properties}\label{sec:vib}

Here, we define all the quantities needed in the calculation of
vibrational properties explicitly; they will be needed in describing
the embedding procedure in Sec.~\ref{sec:embedd}. Central parameters
in these calculations are the Hessian matrix elements:
\begin{equation}
  \Phi_{\alpha,\beta}(m, n) = \frac{\partial F_{m,\alpha}}{\partial r_{n,\beta}},
\label{eq:Hessian}
\end{equation}
where $F_{m,\alpha}$ is the force that acts on atom $m$ in the
Cartesian direction $\alpha$ and $r_{n, \beta}$ is the displacement of
atom $n$ from equilibrium in the direction $\beta$. The Hessian matrix
of the NV center is calculated using the finite-difference
approach~\cite{kresse1995ab}.  This requires a large number of SCF
calculations; however, this number can be reduced employing symmetry
properties. The dynamical matrix element is defined as
$D_{\alpha,\beta}(m,n)=\Phi_{\alpha,\beta}(m,n)/\sqrt{M_\alpha
  M_\beta}$, where $M_{\alpha}$ and $M_{\beta}$ are atomic
masses. Diagonalization of the dynamical matrix yields mass-weighted
normal modes $\boldsymbol{\eta}_k$ and vibrational frequencies
$\omega_k$ of the defect. We then classify the obtained modes
according to the irreducible representation of the $C_{3v}$ group,
{\it i.e.}, $a_1$, $a_2$, or $e$. Actual calculations of dynamical
matrix elements were performed in $4\times 4\times 4$ supercells for
both PBE and HSE.\@

\section{Embedding methodology\label{sec:embedd}}

In Sec.~\ref{sec:exc} we showed that the parameters of the
configuration coordinate diagram related to the optical excitation of
the NV center are converged within 0.01 eV in the
$4 \times 4 \times 4$ supercell. However, while such supercells allow
calculating the general features of the optical lineshapes, they are
not sufficient if our goal is to obtain lineshapes with high accuracy
and high energy resolution~\cite{Alkauskas2014}.

Let us take the $4\times 4\times 4$ supercell as an example.  As will
be explained in Secs.~\ref{a1} and~\ref{jt}, optical lineshapes
reflect lattice relaxation at the defect caused by the optical
transition. In essence, optical spectra reflect the decomposition of
these relaxations in the basis of vibrational modes.  The
lowest-frequency mode at the $\Gamma$ point in the $4\times 4\times 4$
supercell is 35.0 meV for the NV center in the ground state (PBE
result).  The resulting calculations of optical lineshapes can
therefore not contain any contributions of modes with lower
frequencies. In contrast, the experimental luminescence lineshape
clearly shows the contribution of all acoustic phonons down to zero
frequencies~\cite{Davies1976}. Because the relaxations have the
periodicity of the lattice the problem is not resolved by calculating
the phonon spectrum in the entire Brillouin zone of the supercell, as
only $\Gamma$ phonons contribute (see Sec.~\ref{a1} for a more
quantitative discussion).

We solve this issue by the use of the embedding
methodology~\cite{Alkauskas2014}, which enables us to compute lattice
relaxations and vibrational modes for supercells $N \times N \times N$
for which direct first-principles calculations are too expensive. We
apply the methodology for supercells with $N \geqslant 5$. The idea
has two principal components: (i)~calculations of lattice relaxations
in these large supercells; (ii)~calculation of vibrational spectra in
these supercells. At the core of our methodology is the fact that
interatomic interactions in diamond are short-ranged. The meaning of
this statement and the importance for the two aspects mentioned above
are as follows: (a) when the electronic structure of the defect
changes from $\tE$ to $\tA2$ (or vice versa) for \textit{fixed}
lattice positions, the forces that appear on the atoms surrounding the
defect decay fast as a function of the distance from the defect; and
(b) when the position of one atom changes in a fixed electronic
state, the induced force on neighboring atoms also decays very fast as
a function of a distance from this atom. Property (a) enables us to
calculate lattice relaxations in very large supercells.  A
quantitative description will be given in Secs.~\ref{a1}
and~\ref{jt}. Property (b) enables the construction of the Hessian
matrix, and therefore the study of vibrational modes of NV centers
embedded in these large supercells.  The remainder of this Section is
devoted to the explanation of the second component of our
methodology.

The Hessian matrix $\Phi_{\alpha,\beta}(m,n)$ [Eq.~\eqref{eq:Hessian}]
of a large defect supercell is constructed as follows. If atoms $n$
and~$m$ are separated by a distance larger than a chosen cutoff radius
$r_{c1}$, then the Hessian matrix element is set to zero.  If~both
atoms are separated from either the nitrogen atom or the vacant site
by a distance smaller than the cutoff radius $r_{c2}$, then we use the
Hessian matrix element from the actual defect supercell. For all other
atom pairs we use bulk values.

While the procedure is straightforward, it requires some
corrections. Setting matrix elements beyond a certain radius to zero
can break Newton's third law
\begin{equation}
  \label{eq:newton_sum_rule}
   \Phi_{\alpha,\beta}(n,n) = -\sum_{m\neq n}\Phi_{\alpha, \beta}(m, n).
\end{equation}
Breaking this ``acoustic sum rule'' would introduce a small but
nonzero net force on the entire system that could affect the results
for low-frequency acoustic modes. We could enforce the fulfillment of
Newton's third law by setting each matrix element
$\Phi_{\alpha,\beta}(n,n)$ (diagonal in the atomic index $n$) equal to
the rhs of Eq.~\eqref{eq:newton_sum_rule}. However, such a correction
for $\alpha\neq\beta$ can break the symmetry of the Hessian matrix:
\begin{equation}
  \label{eq:symmetry_rule}
  \Phi_{\alpha,\beta}(n,n) = \Phi_{\beta, \alpha}(n, n),
\end{equation}
which follows from a more general symmetry property:
$\Phi_{\alpha,\beta}(m,n) = \Phi_{\beta, \alpha}(n,m)$.  To ensure
that (i) the frequency of acoustic modes at the $\Gamma$ point is
equal to 0 and (ii) symmetry properties of the Hessian are preserved
we set $\Phi_{\alpha,\beta}(n,n)$ to the rhs
of~\eqref{eq:newton_sum_rule} only in the case when $\alpha = \beta$.

Regarding the parameter $r_{c1}$ in the embedding procedure, we used
the value $r_{c1}=7$ \AA.\@ In the case of $r_{c2}$ we used a slightly
smaller value, $r_{c2}=5.6$ \AA, to reduce the overall computational
cost in hybrid functional calculations.  We expect the error bar of
the embedding procedure to be $\sim$2 meV, obtained from the
convergence tests of bulk phonons (see, e.g.,
Ref.~\onlinecite{Londero2018}).

The embedding procedure was applied to supercells up to $N=20$.  A
$N \times N \times N$ supercell contains $8N^3$ atomic sites, and thus
$24N^3$ degrees of freedom. Thus, to find vibrational modes and
frequencies we need to diagonalize dynamical matrices as large as
$192\,000\times 192\,000$. Since these matrices are sparse, with only
$\sim$0.5\% of nonzero elements (sparsity 99.5\%), we used the
spectrum slicing technique~\cite{campos2012strategies} based on the
shift-and-invert Lanczos method, as implemented in the
SLEPc~\cite{Hernandez:2005:SSF} library. Parallelization was done
using an interface to the MUMPS~\cite{amestoy2000multifrontal}
parallel sparse direct solver.

\section{Coupling to \texorpdfstring{$A_1$}{A1} modes\label{a1}}


\subsection{General formulation}

In this Section we discuss the calculation of spectral functions
$A_{a_1} (\hbar\omega)$, as defined in Eqs.~\eqref{a1-lum}
and~\eqref{a1-abs}. The calculation of these spectral functions is
difficult for two main reasons. The first reason is that the
evaluation of overlap integrals
$\braket{\chi^{a_1}_{e0}}{\chi^{a_1}_{gp}}$ or
$\braket{\chi^{a_1}_{es}}{\chi^{a_1}_{g0}}$ entering
equations~\eqref{a1-lum} and~\eqref{a1-abs} is computationally very
challenging, since, generally speaking, normal modes in the ground and
the excited state will not be identical. The two sets of modes are
related via the so-called Duschinsky
transformation~\cite{Osadko1979}. Because the vibrational modes in the
ground and the excited state differ, as explicitly confirmed by our
calculations, overlap integrals are highly multidimensional integrals.
To overcome this problem we will assume the \emph{equal-mode
  approximation}, as is nearly always done for solid-state
systems~\cite{markham1959}.

We will describe the geometry of the entire system using normal
coordinates $Q_k$, i.e.~Cartesian coordinates projected on normal
modes $\boldsymbol{\eta}_{k}$:
\begin{equation}
Q_k = \sum_\alpha \sqrt{M_\alpha} \left(\R_\alpha-\R_{g,\alpha} \right) \boldsymbol{\eta}_{k;\alpha},
\label{eq:normal}
\end{equation}
where $\R_\alpha$ is the position of atom $\alpha$, $\R_{g,\alpha}$ is
its equilibrium position in the ground state, $M_\alpha$ is the mass
of atom $\alpha$, and $\boldsymbol{\eta}_{k;\alpha}$ is a vector that
describes the three components of the mode $\boldsymbol{\eta}_{k}$ for
atom $\alpha$. Within the equal-mode approximation the change of the
adiabatic potential energy surface as a result of optical excitation
is linear in normal coordinates:
\begin{equation}
  \label{eq:HR_pert}
\Delta V (\Q) = V_e(\Q)  - V_g(\Q)  =  \sum_k q_k Q_k,
\end{equation}
where $q_k$ are linear coupling constants. In the expression above we
omit a constant energy offset that does not affect overlap integrals.
In this approximation vibrational modes and frequencies
in the ground and the excited state are identical, but the harmonic
potential describing each vibrational mode $k$ is displaced by
$\Delta Q_k = q_k/\omega_k^2$~\cite{Davies1981}, $\omega_k$ being the
angular frequency of the mode $k$.

The overlap integral of two same-frequency displaced
harmonic-oscillator wavefunctions, pertaining to the vibrational mode
$k$, has an elegant analytical expression~\cite{Davies1981}:
\begin{equation}
  \label{eq:hr_overlap}
  \left|
    \braket{\chi^k_0(Q)}{\chi^k_n(Q - \Delta Q_k)}
  \right|^2 = \frac{S_k^n}{n!}\exp(-S_k)
\end{equation}
where
\begin{equation}
\label{eq:partial}
  S_k = \frac{\omega_k \Delta Q_k^2}{2\hbar}
\end{equation}
is the partial Huang--Rhys (HR) factor.  This factor has a statistical
interpretation as the average number of $k$-mode phonons created
during an optical transition~\cite{Huang1950}.  Partial Huang--Rhys factors define the
so-called spectral density of electron--phonon coupling
$S(\hbar\omega)$, which is the key property that needs to be computed
in order to calculate $A_{a_1} (\hbar\omega)$:
\begin{equation}
  \label{eq:spectralDensity}
  S_{a_1}(\hbar\omega) = \sum_k S_k \delta(\hbar\omega - \hbar\omega_k).
\end{equation}
The total Huang--Rhys factor due to coupling to $a_1$ modes is then
\begin{equation}
S_{a_1} = \int_0^{\infty}S_{a_1}(\hbar\omega) d(\hbar\omega) = \sum_k S_k.
\end{equation}
As seen in the expressions above, we use similar notation for
$S_{a_1}$, the total Huang--Rhys factor and a dimensionless quantity,
and $S_{a_1}(\hbar\omega)$, a spectral density with units
[1/energy]. However, this should cause no confusion, as the spectral
density can always be identified based on the indicated functional
dependence on $\hbar\omega$.

Evaluation of spectral functions~\eqref{a1-lum} and~\eqref{a1-abs} is
simplified if one considers their Fourier transform to the time
domain, commonly denoted as the \textit{generating
  function}~\cite{Lax1952}:
\begin{equation}
  \label{eq:generating-f0}
  G(t) = \int A_{a_1}(\hbar\omega) e^{i\omega t}\,\mathrm{d}(\hbar\omega).
\end{equation}
The generating function for luminescence is given by
\begin{equation}
  \label{eq:generating-lum}
  G(t)
  = \exp\left[
    -iE_{\mathrm{ZPL}}t/\hbar - S_{a_1} + \int e^{i\omega t}S_{a_1}(\hbar \omega)\,\mathrm{d}(\hbar\omega)
  \right],
\end{equation}
and similarly for absorption~\cite{toyozawa2003}:
\begin{equation}
  G(t)
  = \exp\left[
    -iE_{\mathrm{ZPL}}t/\hbar - S_{a_1} + \int e^{-i\omega t}S_{a_1}(\hbar \omega)\,\mathrm{d}(\hbar\omega)
  \right].
  \label{eq:generating-abs}
\end{equation}
Once generating functions are known, spectral functions can be
obtained via the inverse Fourier transform:
\begin{equation}
  A_{a_1}(\hbar \omega) = \frac{1}{2\pi} \int_{-\infty}^{\infty} e^{i\omega t} G(t) e^{-\gamma |t|}
  \, \mathrm{d}t
\label{generating3}
\end{equation}
The term $e^{-\gamma |t|}$ is included to account for the homogeneous
broadening of the optical transition.

The main task is therefore the calculation of partial Huang--Rhys
factors $S_k$ via Eq.~\eqref{eq:partial}. The coefficients
$\Delta Q_k$ in that equation describe the change of the equilibrium
defect geometry upon optical transition and are given by to the
equation similar to Eq.~\eqref{eq:normal}:
\begin{equation}
  \label{eq:q}
   \Delta Q_k = \sum_\alpha \sqrt{M_\alpha} \Delta\R_\alpha \boldsymbol{\eta}_{k;\alpha}.
\end{equation}
Here $\Delta\R_\alpha = \R_{e,\alpha} - \R_{g,\alpha}$ is the change
of the equilibrium position of atom $\alpha$ between the ground and
the excited state.

As discussed in Sec.~\ref{sec:embedd}, actual geometry relaxations
$\Delta\R_{\alpha}$ extend much farther than can be described using
moderate-size supercells used in first-principles calculations.  To
address this problem we note that within the harmonic approximation
$\Delta Q_k$ can alternatively be expressed as
\begin{equation}
  \label{eq:fc_displ}
  \Delta Q_k = \frac{1}{\omega_k^2}\sum_{\alpha}\frac{\F_{\alpha}}{\sqrt{M_{\alpha}}}
  \boldsymbol{\eta}_{k;\alpha},
\end{equation}
where $\F_\alpha$ is the force on atom $\alpha$ induced by the
electronic transition. Specifically, in the case of luminescence this
is the force in the equilibrium geometry of the excited state when the
wavefunction is that of the ground state [point $B$ in
Fig.~\ref{fig:nv}(c)]. In the case of absorption it is the force in
the equilibrium geometry of the ground state when the wavefunction is
that of the excited state [point $D$ in Fig.~\ref{fig:nv}(c)].  In the
case of elastic interactions, as is the case for the NV centre in
diamond, $\F_\alpha$ decays much more rapidly with distance from the
defect center compared to $\Delta\R_\alpha$. Indeed, when the
electronic state is changed, only atoms in the immediate surrounding
experience the change in the force. However, once those atoms start to
move under the influence of these forces, the resulting displacements
$\Delta\R_\alpha$ are long-ranged (see also Appendix~B of
Ref.~\onlinecite{Alkauskas2014}).

We tested that contributions to the spectral density of
electron--phonon coupling $S_{a_1}(\hbar\omega)$ are already converged
if we ignore the forces for atoms lying further than than $r_{c1}=7$
\AA\ from the nitrogen site. Because of this rapid decay of forces,
shown in Sec.~I of the Supplemental Material~\cite{supp}, we make an
approximation that the same forces would be obtained in larger
supercells that are not amenable to explicit first-principles
calculations. This constitutes aspect (i) of the embedding methodology
discussed in Sec.~\ref{sec:embedd}.

Equations~\eqref{eq:partial},~\eqref{eq:spectralDensity},
and~\eqref{eq:fc_displ} define the procedure for the calculation of
$S_{a_1}(\hbar \omega)$ within the equal-mode approximation.  However, since
vibrational modes in the ground and the excited state \textit{are}
different, one has to choose which vibrational modes to use in the
calculation.  In Sec.~II of the Supplemental Material we show, by
means of a simplified model, that in the case of luminescence it is
best to choose vibrational modes and frequencies of the ground state,
while in the case of absorption it is best to choose those of the
excited state. This is the choice we will be making for all the
results reported in this paper.

\subsection{Results: luminescence}


\begin{figure}
  \includegraphics{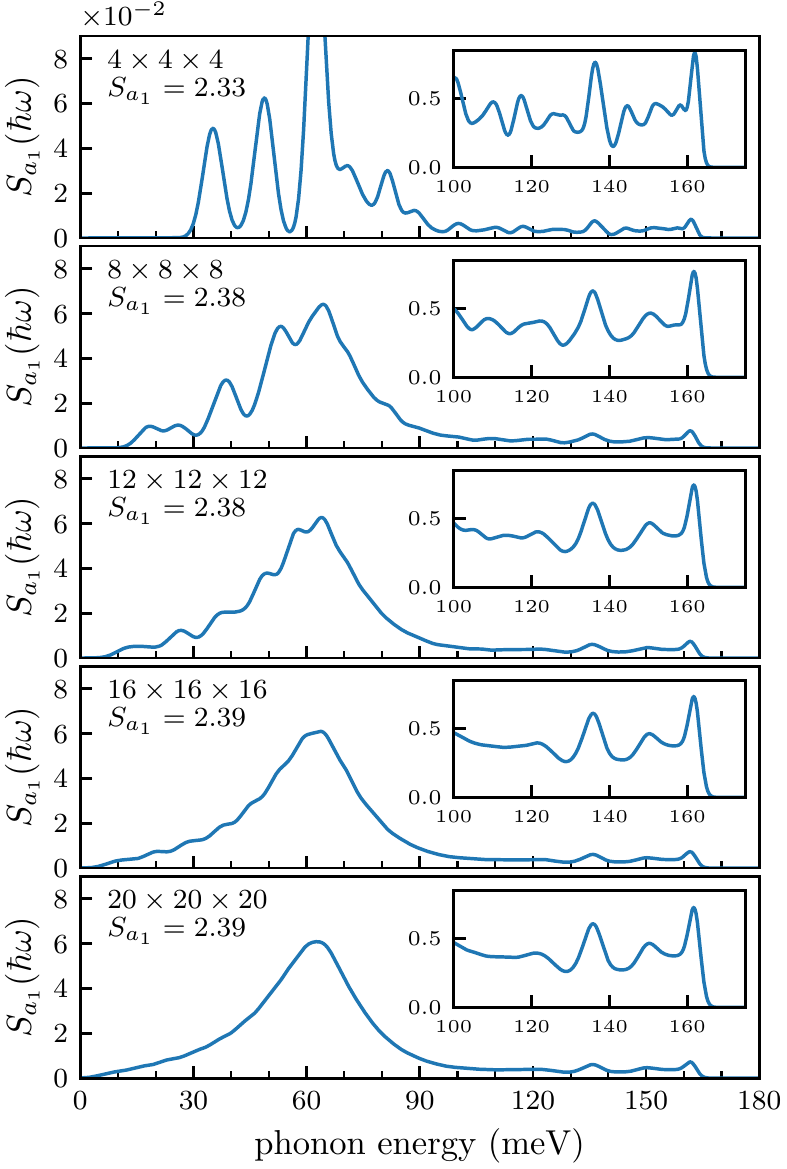}
  \vspace{-1em}
  \caption{
    Convergence of spectral densities $S_{a_1}(\hbar\omega)$ (in units 1/meV) 
    due to coupling to $a_1$ phonons
    with respect to the supercell size.
    Supercells range in size from $4\times 4 \times 4$ (512 atomic sites) to
    $20\times 20 \times 20$ (64\,000 sites).
    Huang--Rhys factors for each supercell are also given.
    The insets enlarge the high-frequency part.
    Gaussian smearing with varying $\sigma$ was used, as explained in the text.\label{fig:convergence}} 
  \vspace{-1em}
\end{figure}

Figure~\ref{fig:convergence} shows how the spectral density due to the
coupling to $a_1$ phonons [Eq.~\eqref{eq:spectralDensity}] converges
as a function of the supercell size. The results pertain to the PBE
functional, and $\delta$-functions in Eq.~\eqref{eq:spectralDensity}
have been replaced by Gaussians. In order to obtain a smooth function
throughout the whole vibrational spectrum we have used Gaussians of
variable width. Our tests indicated that choosing $\sigma$ to vary
linearly from $\sigma=3.5$ meV for $\omega=0$ to $\sigma=1.5$ meV for
the highest-energy phonon results in a smooth spectral density without
introduction of any artifacts.  This smearing procedure will be used
for all spectral densities in this paper.

Figure~\ref{fig:convergence} shows that for phonon frequencies
$>100$~meV the spectral density is already converged for supercells
$12 \times 12 \times 12$ or even $8 \times 8 \times 8$ (inset).
However, larger supercells are needed to converge
$S_{a_1}(\hbar\omega)$ for energies $<60$~meV.  The essence of our
embedding procedure discussed in Sec.~\ref{sec:embedd} was exactly to
achieve this smooth behavior throughout the whole phonon spectrum. It
is also because of a slower convergence in the low-frequency part of
the spectrum that the smearing procedure with a varying $\sigma$ was
used.

\begin{figure}
  \includegraphics[width=8.5cm]{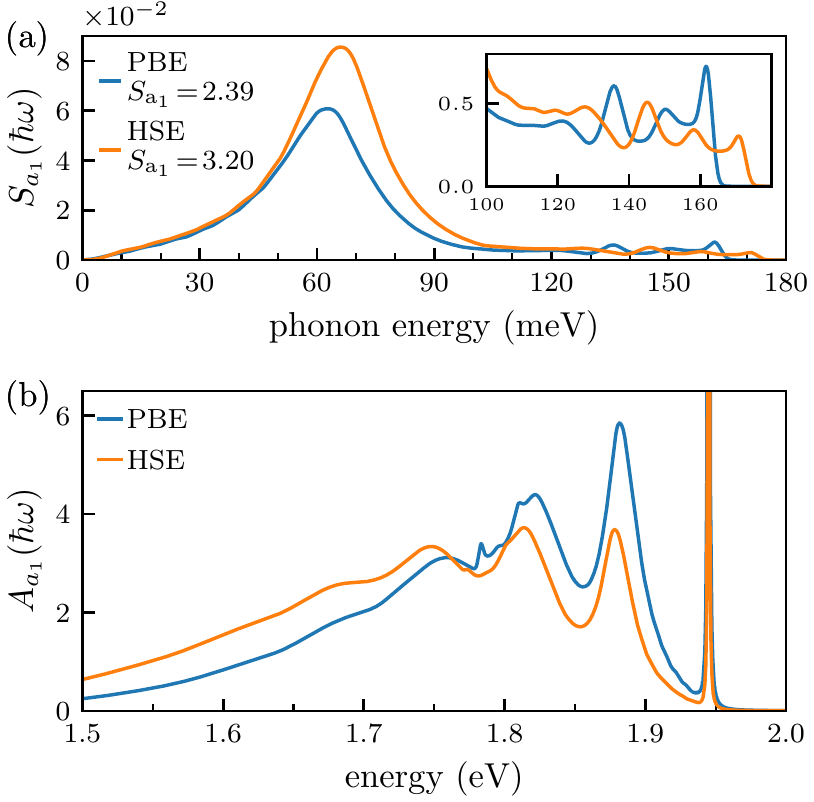}
  \caption{ (a) Spectral densities $S_{a_1}(\hbar\omega)$ (in units
    1/meV) due to coupling to $a_1$ phonons for luminescence,
    calculated with PBE and HSE functionals.  Huang--Rhys factors are
    also given.  The inset enlarges the high-frequency part.  (b)
    Spectral functions $A_{a_1}(\hbar\omega)$ [in units 1/eV,
    Eq.~\eqref{a1-lum}] for luminescence calculated using PBE and
    HSE.\@ The ZPL energy is set to the experimental
    value.\label{fig:a1_lum}}
\end{figure}

PBE and HSE spectral densities for the $20\times20\times20$ supercells
are compared in Fig.~\ref{fig:a1_lum}(a). HSE yields noticeably
stronger electron--phonon interactions with the total HR factor being
$\sim$33 \% larger. The most pronounced features in the HSE spectral
density are shifted to slightly higher phonon energies.  For example,
the ratios $\omega_{\mathrm{HSE}}/\omega_{\mathrm{PBE}}$ for the three
most-pronounced peaks in the PBE spectrum, at 62.5, 135.7 and 161.6
meV, are 1.054, 1.057 and 1.051.  This difference stems from the fact
that the bonds are stiffer in HSE in comparison to PBE, as reflected
in the difference of lattice constants.  In fact, these ratios are
very similar to the ones for bulk phonons given in
Table~\ref{tab:lattice}.  Apart from this, the shapes of the two
spectral densities are rather similar.

Calculated spectral functions $A_{a_1}(\hbar\omega)$
[Eq.~\eqref{a1-lum}] for the two functionals are shown in
Fig.~\ref{fig:a1_lum}(b).  These functions were calculated using
Eq.~\eqref{eq:generating-lum} and $\gamma=0.3$ meV in
Eq.~\eqref{generating3}.  Since the actual luminescence line contains
contributions from both $a_1$ and $e$ phonons, we leave the comparison
with experiment to Sec.~\ref{sec:comb}.


\subsection{Results: absorption\label{sec:a1_abs}}

To calculate $S_{a_1}(\hbar \omega)$ for absorption one needs to
calculate forces in the excited state $\F_\alpha$ at the geometry of
the ground state [point D in Fig.~\ref{fig:nv}(c)], where the two
$^3 E$ states are degenerate.

Due to issues related to the convergence of the electronic structure
for the degenerate state, we calculate these forces indirectly from
displacements via:
\begin{equation}
  \F_\alpha = \sqrt{M_\alpha}\sum_{k}\omega_{k}^{2}\boldsymbol{\eta}_{k;\alpha}\Delta Q_k,
  \label{F_displ}
\end{equation}
where $\omega_k$ and $\boldsymbol{\eta}_{k;\alpha}$ are calculated in
the $4\times 4 \times 4$ supercell, and $\Delta Q_k$ are given by
Eq.~\eqref{eq:q}.  At a first glance, the procedure to calculate the displacements
might appear circular. This is not the case. We use the calculated displacements
in the $4\times 4 \times 4$ supercell to ``restore'' the forces in this supercell.
These forces are then used to determine the displacements in large supercells
for which actual first-principles calculations are not possible. 

\begin{figure}
\includegraphics[width=8.5cm]{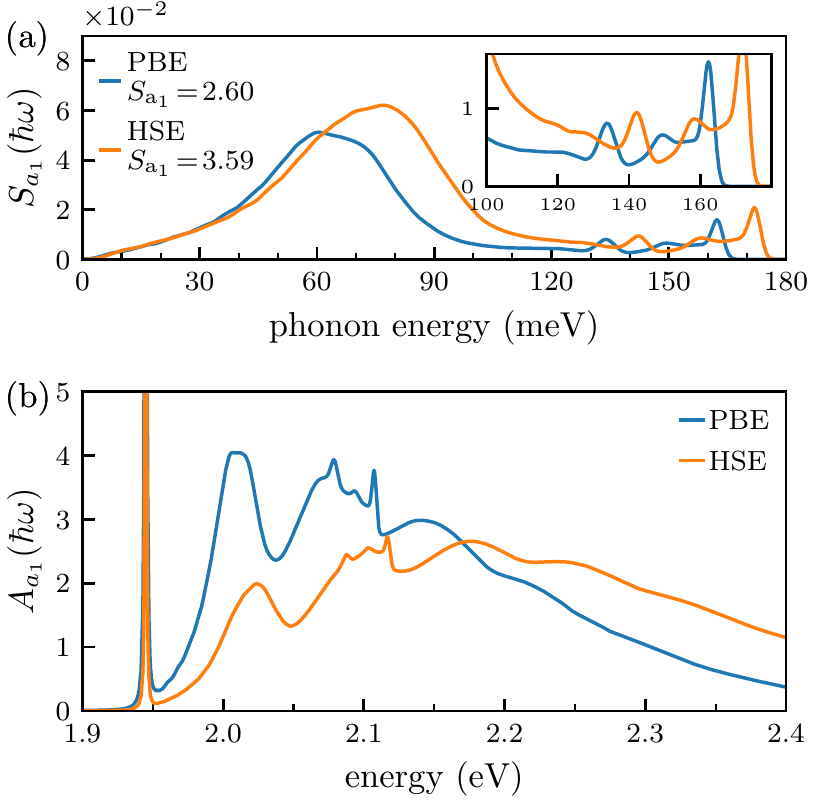}
\caption{ (a) Spectral densities $S_{a_1}(\hbar\omega)$ (in units 1/meV) due to
  coupling to $a_1$ phonons for absorption, calculated with PBE and
  HSE functionals. Huang--Rhys factors are also given.  The inset
  enlarges the high-frequency part.  (b) Spectral functions
  $A_{a_1}(\hbar\omega)$ [in units 1/eV, Eq.~\eqref{a1-abs}] for absorption calculated
  using PBE and HSE.\@ The ZPL energy is set to the experimental
  value.\label{fig:a1_abs}}
\end{figure}

Spectral densities $S_{a_1}(\hbar \omega)$ for absorption in the case
of the $20\times 20 \times 20$ supercell are shown in
Fig.~\ref{fig:a1_abs} (a).  Compared to emission, one sees slightly
larger differences between the shapes of spectral densities calculated
in PBE and HSE.\@ PBE calculations exhibit a broad peak at
${\sim}60$~meV and this peaks shifts to ${\sim}75$~meV in HSE.\@
Spectral functions $A_{a_1}(\hbar\omega)$ for absorption
[Eq.~\eqref{a1-abs}] for the two functionals are shown in
Fig.~\ref{fig:a1_abs} (b).  As in the case of luminescence, the
comparison to experiment is left for Sec.~\ref{sec:comb}.

\section{Coupling to \texorpdfstring{$E$}{E} vibrations: multi-mode
  Jahn--Teller problem\label{jt}}

The coupling to $e$ modes during optical transitions occurs because of
the JT effect in the $\tE$ state. As discussed in Sec.~\ref{sec:exc},
the effect is dynamical.~\cite{Davies1981} The dynamical nature of the
effect has been confirmed by the ${\sim}T^5$ broadening of the ZPL at
low temperatures,\cite{Fu2009} as well as by actual calculations of
the potential energy surface.\cite{abtew2011dynamic}

The purpose of the present Section is to calculate spectral functions
of coupling to $e$ vibrations, given by Eqs.~\eqref{e-lum}
and~\eqref{e-abs}.

\subsection{Multi-mode Jahn--Teller effect: general theory and the
  choice of basis}

The dynamical JT effect is described in a number of articles and
books, e.g., Refs.~\onlinecite{Davies1981,Longuet1958,bersuker2012}.
Here we briefly review the general theory of this effect, and derive
the expressions that are used in our calculations.

The vibronic wavefunction of the excited state is given in
Eq.~\eqref{WF2}.  $a_1$-symmetry vibrational wavefunctions
$\vwf[e][s][a_1][a_1]$ were addressed in Sec.~\ref{a1}. The
$e$-symmetry part
$\ggwf[][et] = \vwf[e][t][e_x][e]\ket{E_x} +
\vwf[e][t][e_y][e]\ket{E_y}$ [Eq.~\eqref{eq:vibronic_term}] can be
represented in the $\{\ket{E_x}, \ket{E_y}\}$ basis as a two-component
vector $ (\chi_{et}^{e_x}, \chi_{et}^{e_y})$.  In this representation
$\ggwf[][et]$ is an eigenvector of the vibronic
Hamiltonian~\cite{bersuker2012}:
\begin{equation}
  \label{eq:vibronic_hamiltonian}
  \Hop = \Hop_{\mathrm{0}} + \Hop_{\mathrm{JT}}.
\end{equation}
Here
\begin{equation}
  \Hop_{\mathrm{0}} = \C{z} \sum_{k;\gamma\in x, y}
  \left(
    -\frac{\hbar^2}{2}\frac{\partial^2}{\partial Q^2_{k\gamma}}
    + \frac{1}{2}  \omega_k^2 Q_{k\gamma}^2
  \right)
\label{eq:Hph}
\end{equation}
describes the motion in the harmonic potential, while
\begin{equation}
  \Hop_{\mathrm{JT}} = \sum_{k;\gamma\in x, y}\C{\gamma} V_k Q_{k\gamma}
\label{eq:JT}
\end{equation}
is the linear Jahn--Teller interaction. $\omega_k$ are angular
frequencies of vibrations, $V_k$ are vibronic coupling coefficients,
and $k=1 \dots N$ runs over all pairs of degenerate $e$-symmetry
vibrations. In the expressions above $\C{\gamma}$ are matrices~\cite{Ham1968}:
\[
  \C{x} =
  \begin{pmatrix}
    0 & 1 \\ 1 & 0
  \end{pmatrix},
  \quad
  \C{y} =
  \begin{pmatrix}
    1 & 0 \\ 0 & -1
  \end{pmatrix},
  \quad
  \C{z} =
  \begin{pmatrix}
    1 & 0 \\ 0 & 1
  \end{pmatrix}.
\]

The Schr\"{o}dinger equation
$\Hop\ggwf[][et]=\varepsilon_{et}\ggwf[][et]$ can be solved by
diagonalizing the Hamiltonian in the basis of eigenvectors of
$\Hop_{\mathrm{0}}$~\cite{bersuker2012}.  One of the obvious choices
for this basis are wavefunctions of the type
$\left |n_{1x}n_{1y} \dots n_{Nx}n_{Ny}; E_x\right \rangle$ and
$\left |n_{1x}n_{1y} \dots n_{Nx}n_{Ny}; E_y\right \rangle$.  $n_{kx}$
and $n_{ky}$ are vibrational quantum numbers pertaining to the
doubly-degenerate mode $k$.  However, this choice of the basis is not
the most convenient when dealing with many $e$ modes. A more
convenient choice is provided by so called ``chiral'' phonons, as
explained in the following.

As a preliminary, let us consider~\cite{Longuet1958} the operator
$\hat{J} = \hat{J}_{\mathrm{el}} + \hat{J}_{\mathrm{ph}}$, where
\begin{align}
  \Jel  = \frac{\hbar}{2}\hat{\sigma}_y, \quad
  \Jph  = \C{z} \sum_k \LL_{z,k}.
\end{align}
Here $\Jph$ is the sum of phonon angular momentum operators
$\LL_{z,k}=i\hbar \left ( Q_{kx}\partial/\partial Q_{ky} - Q_{ky}
  \partial/\partial Q_{kx} \right )$ that acts in a two-dimensional
subspace of normal modes $\{Q_{kx}, Q_{ky}\}$. $\Jel$ acts on the
orbital part of the wavefunction and $\hat{\sigma}_y$ is the Pauli
matrix.  $\hat{J}$ commutes with both $\Hop_{\mathrm{0}}$ and
$\Hop_{\mathrm{JT}}$, and therefore also with $\Hop$.

Let us now find wavefunctions that would be eigenstates of
$\Hop_{\mathrm{0}}$, and simultaneously of $\Jel$ and $\Jph$. The
orbital eigenstates of the operator $\Jel$ are
$\ket{E_{\pm}} = 1/\sqrt{2}(\ket{E_x} \pm i\ket{E_y})$ with
eigenvalues $J_{\mathrm{el}} = \pm \frac{\hbar}{2}$. Expressing
$J_{\mathrm{el}}= j_{\mathrm{el}} \hbar$ enables us to define a
quantum number $j_{\mathrm{el}}=\pm 1/2$, usually called the
electronic pseudo-spin.  Since $\Hop_{\mathrm{0}}$ does not mix
different electronic states, $\ket{E_{\pm}}$ [or
$\left(1/\sqrt{2},\pm i/\sqrt{2}\right)$ in the assumed matrix
notation] are also eigenstates of $\Hop_{\mathrm{0}}$.

To find common eigenstates of $\Jph$ and $\Hop_{\mathrm{0}}$ we will
describe the vibrational degrees of freedom by the aforementioned
``chiral'' phonons. Second-quantization operators of these phonons
are~\cite{cohen1986quantum}:
\[
  a_{k+} = \frac{1}{\sqrt{2}}\left(a_{kx} - ia_{ky} \right), \quad
  a_{k-} = \frac{1}{\sqrt{2}}\left(a_{kx} + ia_{ky} \right),
\]
where $a_{kx}$ and $a_{ky}$ pertain to normal modes $Q_{kx}$ and
$Q_{ky}$, respectively.  Defining the number operator of right- and
left-hand phonons as
\(
  \hat{n}_{k\pm} = a^{\dagger}_{k\pm}a_{k\pm}
\)
we can rewrite the total phonon angular momentum operator as
\begin{equation}
  \Jph = \C{z} \hbar\sum_k(\hat{n}_{k+} - \hat{n}_{k-})
\label{chiral1}
\end{equation}
and $\Hop_{\mathrm{0}}$ as
\begin{equation}
  \Hop_{\mathrm{0}} = \C{z} \sum_k\hbar\omega_k(\hat{n}_{k+} + \hat{n}_{k-} + 1) .
\label{chiral2}
\end{equation}
By comparing Eqs.~\eqref{chiral1} and~\eqref{chiral2} we see that the
common eigenfunctions of $\Hop_{0}$ and $\Jph$ can be
described by two quantum numbers for each pair of $e$ phonons $k$:
$n_k=n_{k+} + n_{k-}$, the total number of $k$-phonons, and
$l_k=n_{k+} - n_{k-}$, whereby $L_{k}=l_k \hbar$ is the angular
momentum quantum number associated with $k$-phonons.  Since $n_{k+}$
and $n_{k-}$ are integers, for a given $n_k$, $l_k$ takes values
$l_k = n_k, n_k-2, n_k - 4, \ldots, -n_k$.

It follows from the previous discussion that common eigenstates of
$\Hop_{\mathrm{0}}$, $\hat{J}_{\mathrm{el}}$, and
$\hat{J}_{\mathrm{ph}}$ can be written as
$\left |n_{1}l_{1} \dots n_{N}l_{N}; E_{+}\right \rangle$ and
$\left |n_{1}l_{1} \dots n_{N}l_{N}; E_{-}\right \rangle$. As they are
also eigenfunctions of $\hat{J}= \Jel + \Jph$, these wavefunctions can
be characterized by a quantum number
\[
  j = j_{\mathrm{el}} + \sum_k l_k.
\]

The new basis functions are eigenfunctions of $\Hop_{0}$, but,
certainly, not of $\Hop_{\mathrm{JT}}$.  If we express $Q_{k\gamma}$ in
the Jahn--Teller Hamiltonian Eq.~\eqref{eq:JT} in terms of creation and
annihilation operators $a_{k\pm}$ and $a_{k\pm}^\dag$, we can rewrite
$\Hop_{\mathrm{JT}}$ as
\begin{equation}
  \label{eq:1}
  \Hop_{\mathrm{JT}} = \sqrt{2}\sum_k K_k\hbar\omega_k
  \begin{pmatrix}
    0 & a_{k+} + a_{k-}^{\dagger}\\
    a_{k-} + a_{k+}^{\dagger} & 0\\
  \end{pmatrix}.
\end{equation}
The parameters $K_k=V_k/\sqrt{2\hbar\omega_k^3}$ are dimensionless
vibronic constants~\cite{obrien1972}. We can then derive matrix
elements of $\Hop_{\mathrm{JT}}$ in the new basis:
\begin{eqnarray}
  && \matrixel{n'_{1}l'_{1},\ldots,n'_{N}l'_{N};E_{-}}
     {\Hop_{\mathrm{JT}}}
     {n_{1}l{}_{1},\ldots,n_{N}l_{N};E_{+}} \notag\\
  && \quad  = \sqrt{2}\sum_{k}K_k\hbar\omega_k\delta_{l'_{k}l_{k}+1}
     \left[\prod_{j\neq k}\delta_{n'_{j}n_{j}}\delta_{l'_{j}l_{j}}\right]
     \notag\\
  && \qquad \times \left[\!\sqrt{\tfrac{n_{k}-l_{k}}{2}}
     \delta_{n'_{k}n_{k}-1}+\sqrt{\tfrac{n_{k}+l_{k}+2}{2}}\delta_{n'_{k}n_{k}+1}\!\right].
     \label{eq:matrix_element}
\end{eqnarray}
We note that $\Hop_{\mathrm{JT}}$ couples only electronic states of
different electron pseudo-spin.  $\Hop_{\mathrm{0}}$ is diagonal in both
vibrational quantum numbers and orbital degrees of freedom with matrix
elements:
\begin{eqnarray}
  && \matrixel{n_{1}l_{1},\ldots,n_{N}l_{N};E_{\pm}}
     {\Hop_{\mathrm{0}}}
     {n_{1}l{}_{1},\ldots,n_{N}l_{N};E_{\pm}} \notag\\
  && \quad  = \sum_k \hbar \omega_k \left ( n_k + 1\right) .
     \label{eq:matrix_element2}
\end{eqnarray}
Equations~\eqref{eq:matrix_element} and~\eqref{eq:matrix_element2} are
the final expressions for the vibronic Hamiltonian in the basis of
``chiral'' phonons.

The logic for choosing the new basis can be recapped as follows.
$\Hop_{\mathrm{JT}}$ and thus
$\Hop=\Hop_{\mathrm{0}}+\Hop_{\mathrm{JT}}$ do not commute with
$\hat{J}_{\mathrm{el}}$ and $\hat{J}_{\mathrm{ph}}$ separately.
However, as discussed above, $\Hop$ commutes with
$\hat{J}_{\mathrm{el}}+\hat{J}_{\mathrm{ph}}$. Therefore the
Hamiltonian only couples basis states with the same quantum
number~$j$, thus separating the diagonalization problem for different
angular momentum components $j$~\cite{Longuet1958}. This is the
biggest advantage of the new basis. The procedure for diagonalizing
$\Hop = \Hop_{\mathrm{0}} + \Hop_{\mathrm{JT}}$ is described in
Sec.~\ref{sec:multimodex}.

\subsection{Calculation of coupling parameters}

The first task in the solution of the vibronic problem is to calculate
linear coupling constants $V_k$ or, alternatively, dimensionless
parameters $K_k$. In the case of the JT effect the adiabatic potential
energy surface in the subspace $\{Q_{kx}, Q_{ky}\}$ (a single $e$
pair) has two branches with energies~\cite{bersuker2012}:
\begin{equation}
  \label{eq:jt_apes}
  U_k(Q_k) = \frac{1}{2}\omega_k^2 Q_k^2 \pm VQ_k,
\end{equation}
where $Q^2_{k}= Q^2_{kx} + Q^2_{ky}$.  Energy minimization in DFT
self-consistent calculations follows the lower-lying branch of
Eq.~\eqref{eq:jt_apes}. The minimum of this potential occurs at
$\Delta Q_{k}=V_k/\omega^2_k$. This enables the determination of $V_k$
and thus $K_k$ once $\Delta Q_{k}$ is known:
\begin{equation}
  K_k^2 = \frac{\omega_k \Delta Q_k^2}{2\hbar}.
\label{eq:kk}
\end{equation}
We find $\Delta Q_k$ from
\begin{equation}
  \Delta Q_k^2 = \Delta Q_{kx}^2 + \Delta Q_{ky}^2,
\label{eq:rho2}
\end{equation}
where
\begin{equation}
  \label{eq:rho}
  \Delta Q_{k\gamma} = \frac{1}{\omega_k^2}\sum_{\alpha}\frac{\F_{\alpha}}{\sqrt{M_{\alpha}}}
  \boldsymbol{\eta}_{k\gamma;\alpha}
\end{equation}
Here, as above, $\gamma=\{x,y\}$. Equation~\eqref{eq:rho} is identical
to Eq.~\eqref{eq:fc_displ} (see Sec.~\ref{a1} for the description of
parameters), with the only difference that Eq.~\eqref{eq:rho} is
applied to $e$ modes rather than $a_1$ modes. In the case of
absorption forces $\F_{\alpha}$ are obtained as in
Sec.~\ref{sec:a1_abs}.

\begin{figure}
\includegraphics[width=8.5cm]{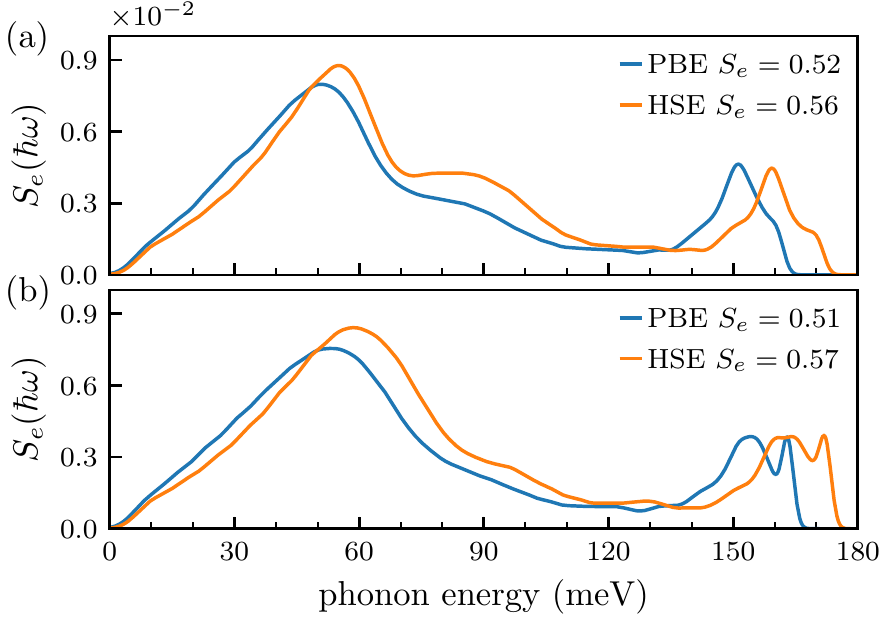}
\vspace{-1em}
\caption{Spectral density $S_e(\hbar\omega)$ (in units 1/meV) due to coupling to $e$
  phonons for (a) luminescence and (b) absorption, calculated with PBE
  and HSE functionals. Results are for the $20 \times 20 \times 20$
  supercell.  Huang--Rhys factors are also given.\label{fig:k2}}
\end{figure}

Comparing Eq.~\eqref{eq:kk} with Eq.~\eqref{eq:partial} we see that
the dimensionless parameter $K_k^2$ plays a similar role to the
partial Huang--Rhys factor $S_k$ in the case of coupling to $a_1$
modes.  A different notation is used for historical reasons.  Thus, in
analogy to Eq.~\eqref{eq:spectralDensity}, we introduce a spectral
density of coupling to asymmetric $e$ modes, given by:
\[
S_e(\hbar\omega) = \sum_k K_k^2\delta(\hbar\omega - \hbar\omega_k),
\]
where the sum runs over all $e$ doublets. As in the case of
$S_{a_1}(\hbar\omega)$, $S_e(\hbar\omega)$ depends on the supercell
size, converging towards the final result as the system size
increases. The result for the $20 \times 20 \times 20$ supercell is
shown in Fig.~\ref{fig:k2}. The integral of this function quantifies
the strength of the Jahn--Teller interaction in a system, which we
will label~$S_e$:
\begin{equation}
S_e = \sum_k K_k^2 = \int_0^{\infty} S_e(\hbar\omega) d(\hbar\omega) .
\label{total:JT}
\end{equation}
For simplicity of the nomenclature we will also call it a Huang--Rhys
factor pertaining to coupling to $e$ modes. Jahn--Teller interaction
is considered strong for $S_e \gg 1$, and weak in the case of
$S_e \ll 1$~\cite{bersuker2012}. For the NV center HSE values are
$S_e = 0.56$ for emission and $S_e = 0.57$ for absorption. One can
conclude that the JT coupling is of medium strength.

\subsection{Luminescence and absorption processes}

Diagonalization of $\Hop = \Hop_{\mathrm{0}} + \Hop_{\mathrm{JT}}$
[Eqs.~\eqref{eq:matrix_element} and~\eqref{eq:matrix_element2}]
produces vibronic wavefunctions~\eqref{eq:vibronic_term} in the form
$ \ket{\Phi_{et}} =
\ket{\vwf*[e][t][+]}\ket{E_+}+\ket{\vwf*[e][t][-]}\ket{E_-} $
where:
\begin{equation} \ket{\vwf*[e][t][\pm]} =
\sum_{\boldsymbol{nl}}C_{et;n_{1}l_{1}\ldots n_{N}l_{n}}^{\pm}
\ket{n_{1}l_{1},\ldots,n_{N}l_{N}}.
\end{equation}
In terms of the wavefunctions $\ket{\vwf*[e][t][\pm]}$ the overlap
integrals that appear in the expression for $A_{e}(\hbar\omega)$ in
Eqs.~\eqref{e-lum} and~\eqref{e-abs} can be rewritten as
\begin{align*}
  \left|\braket{\vwf*[g][r][e][e]}{\vwf*[e][t][e_x][e]}\right|^2 +
  \left|\braket{\vwf*[g][r][e][e]}{\vwf*[e][t][e_y][e]}\right|^2
  =   \left|\braket{\vwf*[g][r][e][e]}{\vwf*[e][t][+][e]}\right|^2 +
  \left|\braket{\vwf*[g][r][e][e]}{\vwf*[e][t][-][e]}\right|^2 .
\end{align*}

In the limit of zero temperature overlaps for luminescence
spectrum~\eqref{e-lum} are calculated between the lowest vibronic
state of the electronic excited state $\ket{^3E}$ and all the
vibrational states of the electronic ground state $\ket{^3A_2}$. The
lowest vibronic state is always the one with the ``pseudo spin''
$j=\pm \tfrac{1}{2}$~\cite{bersuker2012}.  Therefore, in the case of
luminescence one has to diagonalize the Hamiltonian for either the
$j=\tfrac{1}{2}$ or the $j=- \tfrac{1}{2}$ ``channel'' (the two states
are degenerate).

In the case of absorption, overlap integrals in the spectral
function~\eqref{e-abs} are calculated between zero-phonon state
$\ket{00..0}$ of the electronic ground state $\ket{^3A_2}$ and
vibronic states of the electronic excited state $\ket{^3E}$.  Overlaps
will be non-zero only for vibronic states in the $\ket{^3E}$ manifold
that contain the contribution of the zero-phonon state
$\ket{00..0}$. This phonon state is only present in vibronic states
with $j = \pm\tfrac{1}{2}$~\cite{Longuet1958}, and therefore in our
diagonalization procedure we again need to consider either only the
$j=\tfrac{1}{2}$ or the $j=- \tfrac{1}{2}$ ``channel''.

As in the case of spectral functions $A_{a_1}(\hbar\omega)$, when
calculating $A_{e}(\hbar\omega)$ we choose the vibrational modes and
frequencies of the ground state for luminescence, and those of the
excited state for absorption.

\subsection{Diagonalization of the vibronic
  Hamiltonian\label{sec:multimodex}}

Reformulation of the problem in terms of the new basis makes the
diagonalization of the vibronic Hamiltonian in the presence of a
\textit{small} number of $e$ modes a computationally tractable task.
Without this reformulation, treating even a few modes would be
computationally too expensive.  In constructing the basis set we limit
the total number of excited vibrations $n_{\text{tot}} = \sum_k n_k$
to a certain number. By increasing this number we can monitor the
convergence of the final result. We find that convergence is easily
achieved in our case.

However, we are still facing a daunting challenge: for the NV center
effectively an infinite number of $e$ modes, described by the spectral
density $S_e(\hbar\omega)$, participate in the Jahn--Teller effect. To
address this problem we propose the following approach.

We approximate the actual spectral density $S_{e}(\hbar\omega)$ with
$S^{(\mathrm{eff})}_{e}(\hbar\omega)$, defined as:
\begin{equation}
  S^{(\mathrm{eff})}_{e}(\hbar\omega) = \sum_{n=1}^{N_{\mathrm{eff}}}
  \bar{K}^2_{n}g_{\sigma}(\hbar\omega_{n}-\hbar\omega),
\label{eq:Seff}
\end{equation}
Here $g_{\sigma}$ is a Gaussian function of width $\sigma$; the sum
runs over $N_{\mathrm{eff}}$ ``effective'' vibrations with frequencies
$\omega_n$ and vibronic coupling strengths $\bar{K}^2_{n}$. For a
fixed number $N_{\mathrm{eff}}$, the parameters $\bar{K}^2_n$,
$\omega_{n}$, and $\sigma$ are obtained by the minimization of the
integral
\begin{equation}
  I = \int_0^\infty \left|S_e(\hbar\omega) -
    S^{(\mathrm{eff})}_e(\hbar\omega)\right|\,\mathrm{d}(\hbar\omega)
\label{eq:min}
\end{equation}
while enforcing that
$\sum_{n=1}^{N_{\text{eff}}}\bar{K}^2_{n}=\sum_{k=1}^N K_k^2=S_e$.  If
$N_{\mathrm{eff}} = N$, the actual number of $e$ doublets, then this
approach reproduces the full calculation. However, one expects that
convergence of the final result can be achieved for
$N_{\mathrm{eff}}\ll N$, such that the problem is still tractable by
the diagonalization procedure. This indeed turns out to be the case.
In Sec.~III of the Supplemental Material we present a test of the
procedure for the $2\times 2\times 2$ supercell.

\subsection{Spectral functions $A_{e}(\hbar \omega)$ for absorption
  and emission}

In this Section we briefly discuss the spectral functions
$A_{e}(\hbar\omega)$ [Eqs.~\eqref{e-lum} and~\eqref{e-abs}] for both
luminescence and absorption.

\begin{figure}
\includegraphics{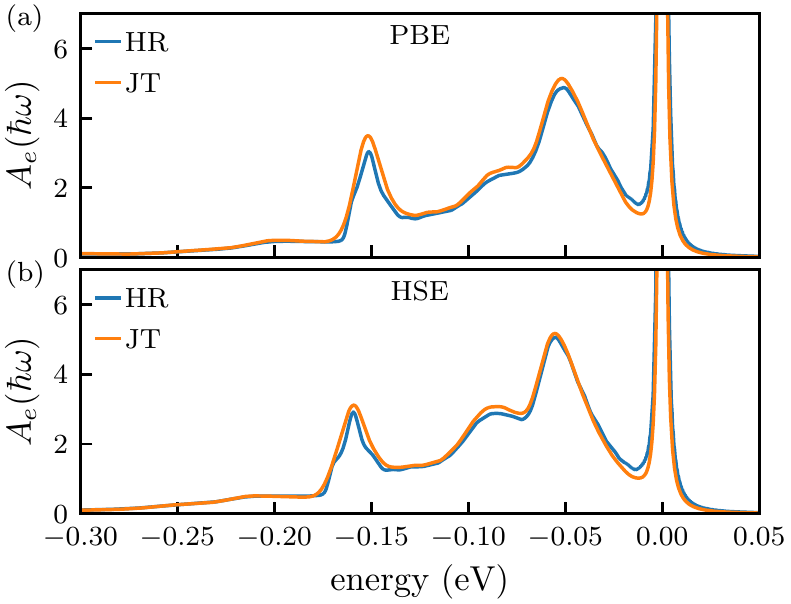}
\caption{The spectral function $A_e(\hbar\omega)$ (in units 1/eV) for
  emission [Eq.~\eqref{e-lum}] obtained with (a) PBE and (b) HSE
  functionals. We compare spectral functions obtained via the solution
  of the multi-mode Jahn--Teller problem (``JT'') and via the
  Huang--Rhys treatment (``HR'').\label{fig:Ae-res-lum}}
\end{figure}

The results for luminescence are shown in Fig.~\ref{fig:Ae-res-lum}
for both functionals.  We used $N_{\mathrm{eff}}=22$
(cf.~Sec.~\ref{sec:multimodex}).  In the figure we compare
$A_{e}(\hbar\omega)$ obtained via the solution of the multi-mode JT
problem (labeled ``JT'') with the one obtained via the Huang--Rhys
approach (labeled ``HR''). In the latter we treat the $e$ modes as if
they were fully symmetric $a_1$ modes. Spectral functions are obtained
in a manner completely identical to those of $a_1$ modes, as described
in Sec.~\ref{a1}.

Somewhat surprisingly, the HR calculation yields spectral functions
very similar to those of the JT calculation. The agreement between the
two sets of calculations is striking and one is tempted to conclude
that there is a deeper underlying reason for this.  However, in
Appendix~\ref{sec:apJT} we show that the good agreement is to some
degree accidental. Indeed, using first-order perturbation theory we
derive that for $S_e\ll 1$ the JT theory yields an intensity for the
first vibrational side-peak exactly twice as large as the HR
approach. As a result, the weight of the ZPL is \textit{smaller} in
the JT treatment. We then use a simple model to show that as $S_e$
becomes larger, the situation is reversed: the weight of the first
phonon peak becomes smaller in the JT calculation with respect to the
HR result, and vice versa for the weight of the ZPL.\@ Interestingly,
in the range $S_e \approx 0.5-1.0$ the two approaches provide a very
similar quantitative description, explaining the result shown in
Fig.~\ref{fig:Ae-res-lum}.  Regardless, our analysis warns that
applying the Huang--Rhys approach can lead to errors in the case of
small $S_e$.

\begin{figure}
\includegraphics{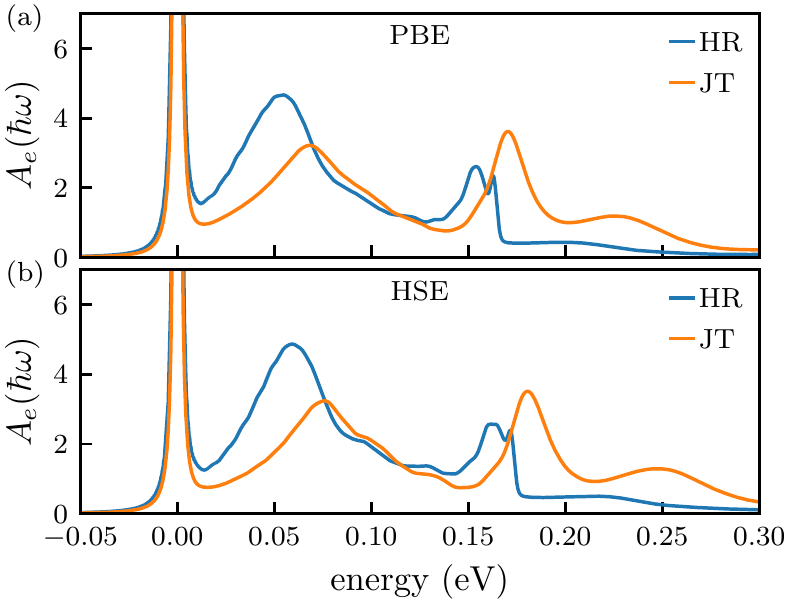}
\caption{The spectral function $A_e(\hbar\omega)$ (in units 1/eV) for absorption
  [Eq.~\eqref{e-abs}] obtained with (a) PBE and (b) HSE
  functionals. We compare spectral functions obtained via the solution
  of the multi-mode Jahn--Teller problem (``JT theory'') and via the
  Huang--Rhys treatment (``HR theory'').\label{fig:Ae-res-abs}}
\end{figure}

The spectral functions $A_e(\hbar\omega)$ for absorption are shown in
Fig.~\ref{fig:Ae-res-abs}. In contrast to luminescence, the
Jahn--Teller treatment differs substantially from the Huang--Rhys
calculation. Overall, compared to the HR function, the JT function is
``stretched''. This is in agreement with model calculations for systems 
with the dynamic JT effect \cite{Longuet1958}, as also exemplified
in recent first-principles modelling of diamondoids~\cite{gali2016}. 
Moreover, we observe a change in the energy and the
intensity of peaks that appear in $A_e(\hbar\omega)$. For example, the
phonon side peak closest to the ZPL decreases in intensity and moves
to larger energies. These results clearly illustrate that using the JT
theory is essential in the case of absorption at NV centers in
diamond.

\section{Results: \texorpdfstring{$A_1$}{A1} and
  \texorpdfstring{$E$}{E} modes combined\label{sec:comb}}

In this Section we present the final result of our calculated,
luminescence and absorption lineshapes, obtained via
Eqs.~\eqref{conv},~\eqref{Lem}, and~\eqref{Lab}, and compare them with
experimental lineshapes~\cite{Manson2018}. In Table~\ref{tab:HR_tot}
the calculated Huang--Rhys factors for the coupling with $a_1$ and $e$
modes are summarized. We define the total Huang--Rhys factor as
$S_{\text{tot}}=S_{a_1}+S_e$. In comparison with experiment, the total
Huang--Rhys factor for emission is slightly underestimated in PBE and
overestimated in HSE.\@ The contribution of $e$ modes to optical
lineshapes can be quantified by a ratio $S_e/S_{\text{tot}}$, which we
find to be $14-18$ \%.

\begin{table}[t]
  \caption{Calculated Huang--Rhys factors for emission and absorption.\label{tab:HR_tot}}
  \begin{ruledtabular}
    \begin{tabular}{lcccccc}
            &\multicolumn{3}{c}{Luminescence} & \multicolumn{3}{c}{Absorption} \\
            \cline{2-4}  \cline{5-7}
            & $S_{a_1}$ & $S_e$ & $S_{\text{tot}}$ & $S_{a_1}$ & $S_e$ & $S_{\text{tot}}$ \\
      \hline
      \\[-1.5ex]
       PBE     & 2.39  & 0.52  &  2.91  & 2.60  & 0.51  &  3.11 \\
       HSE     & 3.20  & 0.56  &  3.76  & 3.59  & 0.57  &  4.16 \\
       expt.   &       &       &  \hspace{1ex}3.49\footnotemark[1]  &       &       &       \\
    \end{tabular}
  \end{ruledtabular}
\footnotetext[1]{Reference \onlinecite{Kehayias2013}}
\end{table}

\subsection{Luminescence}

\begin{figure}
  \includegraphics[width=8.5cm]{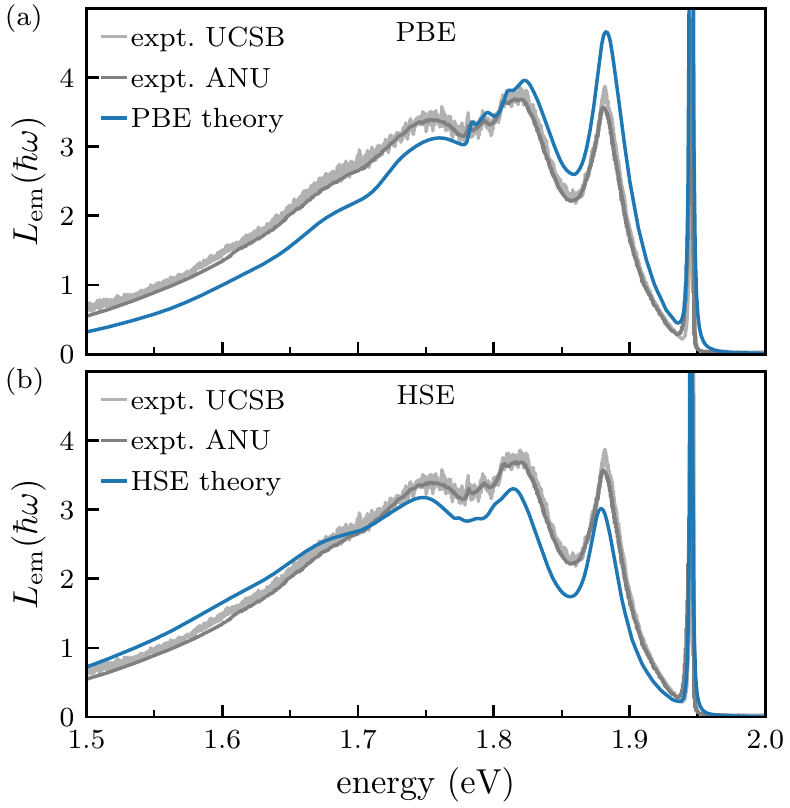}
  \caption{Theoretical normalized luminescence lineshapes (in units 1/eV), compared with
    experimental spectra: (a) PBE functional; (b) HSE
    functional. Experimental spectra from
    Refs.~\onlinecite{Kehayias2013} (labeled ``ANU'') and
    \onlinecite{Alkauskas2014} (labeled ``UCSB''). The ZPL energy of
    the theoretical curves is set to the experimental value.\label{fig:theory_lum}}
\end{figure}

The calculated luminescence lineshapes are compared to experimental
curves from Refs.~\onlinecite{Kehayias2013} (labeled ``ANU'') and
\onlinecite{Alkauskas2014} (labeled ``UCSB'') in
Fig.~\ref{fig:theory_lum}. Details about experimental procedures and
samples are given in the corresponding papers.  To allow for a
meaningful comparison the theoretical lineshapes were shifted to match
the experimental ZPL.\@ The overall agreement is quite good with both
the PBE and the HSE functional. As one of the goals of the current
paper is the analysis of the accuracy of modern density functionals in
describing these luminescence lineshapes, we now discuss the
differences between the two sets of calculations.

Comparing experiment with the PBE calculation, we see that the
intensities of the ZPL and the first two phonon peaks in the
calculated lineshape are too large.  This is because the total HR
factor calculated in PBE is smaller than the experimental one
(Table~\ref{tab:HR_tot}). In contrast, the total HR factor calculated
in HSE is larger than the experimental one (Table~\ref{tab:HR_tot}),
yielding the intensity of the ZPL and the three phonon side peaks too
small in comparison with experiment. From the atomistic point of view,
one could conclude that the change in the equilibrium defect geometry
between the ground and the excited states is slightly underestimated
in PBE, while this change is overestimated in HSE.\@

Looking at the position of the peaks, we see, however, that theory
agrees with experiment remarkably well. This concerns not only the
main phonon replica at about 65 meV, but even the fine structure of
the luminescence lineshape, especially visible between the second and
the third replica of the 65 meV peak.  We do note that the positions
of the peaks in PBE (which are directly related to the vibrational
frequencies) show a very close agreement with experiment, while some
shifts are evident for HSE, because frequencies are slightly
overestimated.  This conclusion is in line with the fact that PBE does
a better job at describing the lattice constant, bulk modulus, and
bulk phonons of diamond (Sec.~\ref{sec:first-princ}).

\subsection{Absorption}

\begin{figure}
  \includegraphics[width=8.5cm]{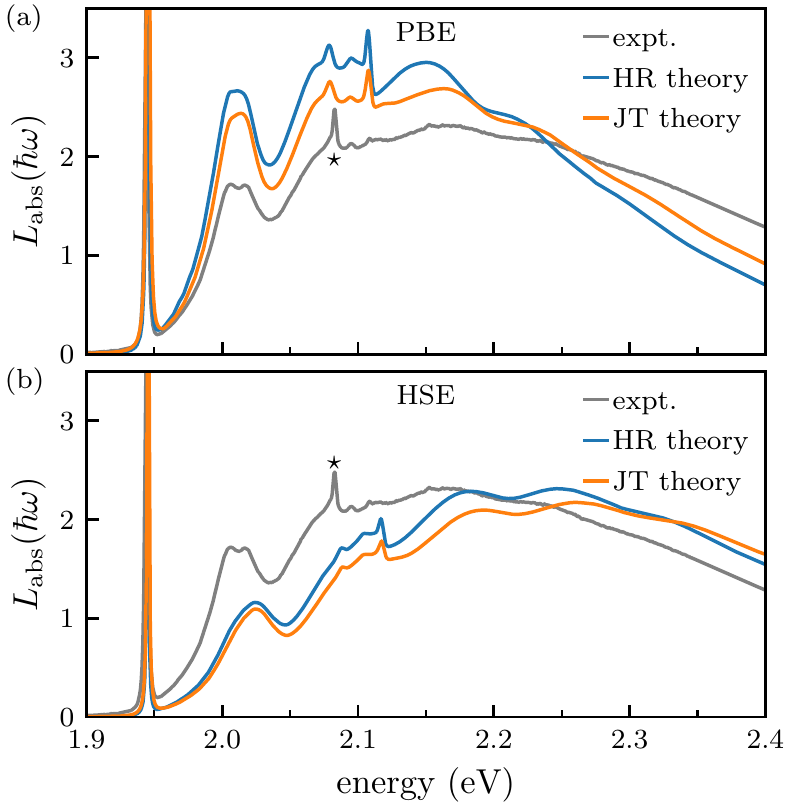}
  \caption{Theoretical normalized absorption lineshapes (in units
    1/eV) calculated using the Huang--Rhys (labeled `HR theory') and
    Jahn--Teller treatment (labeled ``JT theory''), compared with the
    experimental spectrum: (a) PBE functional; (b) HSE functional. The
    experiment is from Ref.~\onlinecite{Manson2018}. The ZPL energy of
    the theoretical curves is set to the experimental value. The small
    peak marked with a star ``$\star$'' in the experimental curve is
    the ZPL of another center and should be disregarded in the
    comparison.\label{fig:theory_abs}}
\end{figure}

The calculated absorption lineshapes, where the contributions of both
$a_1$ and $e$ phonons are included, are compared to experiment in
Fig.~\ref{fig:theory_abs}. The experimental lineshape is from
Ref.~\onlinecite{Manson2018}.  As we did in the case of luminescence,
we will first compare the overall lineshape and then look at the fine
structure of the absorption band.

Comparing the PBE calculation [Fig.~\eqref{fig:theory_abs}(a)] with
experiment we see that, like for luminescence, the intensity of both
the ZPL and the first phonon side peak is overestimated in PBE.\@ In
contrast, both of these intensities are underestimated in HSE
[Fig.~\eqref{fig:theory_abs}(b)].  Like in the case of luminescence,
we trace this to the accuracy of the functionals in describing the
change of the defect geometry in the ground state with respect to that
of the excited state: the lattice relaxation is slightly
underestimated in PBE, and overestimated in HSE.\@

Looking at the fine structure of the spectra, we see that, also on par
with luminescence, the positions of the peaks are better described in
PBE in comparison to HSE.\@ This distinction is particularly clear in
the description of the first phonon side peak, at $\sim$2.1 eV. The
experimental curve displays the famous double-peak
structure~\cite{Davies1976}.  While the double peak is not clearly
revealed in the PBE calculation, one can nevertheless see that the
calculation accurately describes the overall position of the peak. In
addition, the peak is broader than in the case of luminescence, in
line with experimental findings. This peak is shifted to slightly larger energies in HSE.\@ 
This can again be attributed to the fact that, like
for bulk diamond and the NV center in the $\tA2$ state, the
vibrational modes and frequencies in the $\tE$ state are more
accurately described in PBE.\@ The origin of the double-peak structure
will be addressed in a separate experimental--theoretical paper.

Focusing on the PBE result, it is interesting to compare the JT
treatment with the HR treatment. Strictly speaking, the HR treatment
for absorption is not justified in the presence of the dynamical JT
effect. However, it can be viewed as an approximation, and is
computationally much simpler. While the positions of peaks are
reasonably well described in both treatments
[Fig.~\ref{fig:theory_abs}(a)], there are distinct differences.  For
instance, the JT approach offers a better description of the two
features at energies ${\sim}2.15$ and $2.24$~eV.  The improved
agreement with experiment lends support to the validity of the
multi-mode JT approach developed in our work.

\subsection{Summary of comparison with experiment}

The main conclusions of our calculations, discussed in this Section, are:

(i) The change of the equilibrium defect geometry of the NV center
  in the $\tE$ state with respect to the $\tA2$ state is overestimated
  in the HSE functional, while it is underestimated in the PBE
  functional.

(ii) Overall, both functionals describe the major peaks, as well as
  the fine structure in luminescence quite accurately. The positions
  of the peaks in the luminescence band calculated with the PBE
  functional are in nearly perfect agreement with the experimental
  spectrum, while the peak positions are overestimated (relative to
  the ZPL) in HSE.

(iii) The PBE functional describes the positions of major peaks and
  the fine structure in the absorption spectrum better than the HSE
  functional, and this is in particular visible in the width of the
  first phonon side peak.

  (iv) Focusing on the PBE result, the Jahn--Teller theory provides a
  more accurate description of the peak positions in absorption, which
  is especially true in the case of the broad structure in the
  $2.1\mbox{--}2.3$~eV range.

(v) The splitting of the first phonon side-peak in absorption is not
  clearly reproduced in theoretical calculations.

\section{Discussion\label{sec:disc}}

In this Section we critically review our methodology and the
calculations, with a special emphasis on the accuracy of density
functionals in the quantitative description of the vibrational and
vibronic structure of isolated NV centers. Before we discuss the
origin of the remaining discrepancies between theoretical and
experimental curves (Figs.~\ref{fig:theory_lum}
and~\ref{fig:theory_abs}), let us mention some aspects that
\textit{have not} been addressed or that have only been partially
addressed in our current paper:

\emph{Quadratic interactions}. Our calculations rely on the linear
theory of electron--phonon coupling. While there is little evidence
that quadratic terms are important in the coupling to $a_1$ phonons,
it is known that quadratic Jahn--Teller terms do manifest
themselves~\cite{abtew2011dynamic,thiering2017ab}. In particular, it
is estimated that the quadratic terms are ${\sim}1/3$ of the linear
terms in terms of
energy~\cite{abtew2011dynamic,thiering2017ab}. Unfortunately,
inclusion of quadratic terms in the multi-mode treatment increases the
complexity of an already complex problem tremendously and it may not
be possible to include this given current computational
capabilities. We do suggest that more work is needed here.

\emph{Treatment of the excited state}. The
$\tE$ excited state has been described using the
$\Delta$SCF approach, as customary in state-of-the-art defect
calculations~\cite{Gali2009,dreyer2018}. As~demonstrated by our
results,
$\Delta$SCF yields good results, even though the calculation of
excited states using constrained orbital occupations in DFT does not
have the same fundamental backing of rigorous theorems as the
calculation of the ground state. Future work in this field would
certainly benefit from the ongoing developments in calculating excited
states using more rigorous (and computationally much more intensive)
many-body approaches.

\emph{Hessian matrices for charged defects}. There is a remaining
question regarding the accuracy of Hessian matrix elements for charged
defects in actual supercell calculations. In an infinite solid the
negatively charged defect would induce polarization charge
$+ (1-{1/\varepsilon})$ in the vicinity ($\varepsilon$ being the
dielectric constant), while the polarization charge of the same
magnitude but opposite polarity would be pushed to infinity. In the
supercell approach the polarization charge that should be pushed to
infinity is homogeneously distributed over the
supercell~\cite{freysoldt2011} and there is, in addition, a
neutralizing background that is introduced to prevent the Coulomb
interaction from diverging.  This ``unphysical'' spreading of the
polarization charge and the presence of the neutralizing background
may introduce errors in the calculation of force constants. Since
these errors decrease as the size of the supercell grows, the approach
employed in this work is to calculate these constants for the largest
system that is tractable for both PBE and HSE functionals, {\it i.e.},
the $4\times4\times4$ supercell nominally containing 512 items. A~more
rigorous solution of the issue is left for future work.

\emph{Treatment of degenerate electronic states and dynamical Jahn-Teller effect.} As has been discussed in the literature, the application of DFT to degenerate electronic states ($\tE$ in our case) is in principle more troublesome than in the case of non-degenerate ones \cite{bersuker1997}, as the standard Hohenberg-Kohn theorem does not strictly apply. This also translates into practical aspects of finding vibrational frequencies and vibronic coupling constants. In this paper we chose specific approximations to determine these quantities, i.e., determining the vibrational frequencies using the ``symmetric'' electron configuration \sym  and determining vibronic coupling constants from calculations performed away from the degeneracy point.

Let us assume that the aspects mentioned above affect the
calculated lineshapes in a minor way. In this case one could attribute
the remaining discrepancy between experiment and theory to the
accuracy of density functionals in describing structural and
vibrational properties of diamond and the NV center.

To test this hypothesis, let us assume that the shape of the
calculated spectral densities $S_{a_1} (\hbar\omega)$ and
$S_{e} (\hbar\omega)$, calculated in PBE, is close to the ``truth'',
keeping in mind an excellent agreement regarding the positions of
peaks, especially for luminescence. However, as discussed in
Sec.~\ref{sec:comb}, atomic relaxations are slightly underestimated in
PBE, the conclusion reached comparing the calculated Huang--Rhys
factor with the experimental one.  Let us assume that atomic
relaxations projected on all vibrational modes are underestimated by
the same linear factor $\zeta^{1/2}$. As a result, our estimate for
``corrected'' spectral densities are
$S'_{a_1} (\hbar\omega)=\zeta S_{a_1} (\hbar\omega)$ and
$S'_{e} (\hbar\omega)=\zeta S_{e} (\hbar\omega)$.

\begin{figure}
  \includegraphics[width=8.5cm]{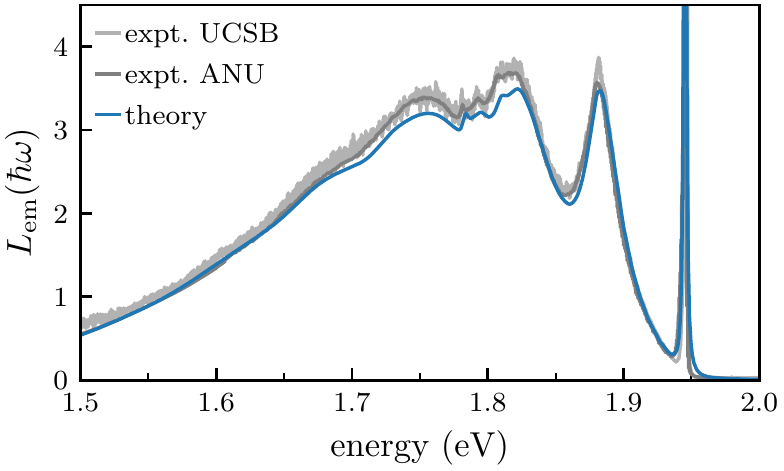}
  \caption{Calculated normalized luminescence lineshape, compared with
    the experimental lineshape (both in units 1/eV). In the
    calculations we used scaled PBE spectral densities
    $S'_{a_1}=\zeta S_{a_1} (\hbar\omega)$ and
    $S'_{e}=\zeta S_{e} (\hbar\omega)$, with $\zeta=1.2$. Experimental
    spectra as in Fig.~\ref{fig:theory_lum}.\label{fig:lum_corr}}
\end{figure}

In Fig.~\ref{fig:lum_corr} we show the calculated luminescence
lineshape with $\zeta=1.2$; $\zeta$ was obtained via a least-square
fit to the experimental luminescence lineshape. Unsurprisingly, this
value is almost exactly the ratio between the experimental total
Huang--Rhys factor and the PBE value ($3.49/2.91=1.199$;
cf.~Table~\ref{tab:HR_tot}). It is evident that the agreement with
experiment is very good, both for the general shape of the
luminescence line and for all the fine features of the spectrum.

\begin{figure}
  \includegraphics[width=8.5cm]{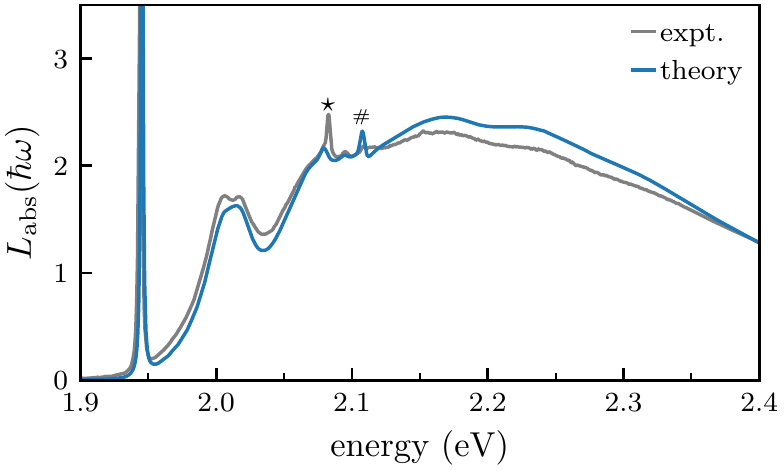}
  \caption{(Color online) Calculated normalized absorption lineshape,
    compared with the experimental lineshape (both in units 1/eV). In
    the calculations we used scaled PBE spectral densities
    $S'_{a_1}=\zeta S_{a_1} (\hbar\omega)$ and
    $S'_{e}=\zeta S_{e} (\hbar\omega)$, with $\zeta=1.2$. Experimental
    spectrum as in Fig.~\ref{fig:theory_abs}. The experimental peak
    marked ``$\star$'' is the ZPL of another defect and should be
    disregarded in the comparison.  The feature marked ``$\#$'' in the
    calculated curve is discussed in the text.\label{fig:abs_corr}}
\end{figure}

Since the value of $\zeta$ represents the scaling of the geometry
relaxation in the excited state with respect to the ground state, we
have to use the same value of $\zeta$ for the calculation of the
absorption lineshape. In other words, once $\zeta$ is fixed based on
an analysis of the luminescence lineshape, there are no more free
parameters in the calculation of the absorption lineshape. Like for
luminescence, in this calculation we use PBE spectral densities
$S_{a_1} (\hbar\omega)$ and $S_{e} (\hbar\omega)$ scaled with $\zeta$:
$S'_{a_1} (\hbar\omega)=\zeta S_{a_1} (\hbar\omega)$ and
$S'_{e} (\hbar\omega)=\zeta S_{e} (\hbar\omega)$. The result of the
calculation is shown in Fig.~\ref{fig:abs_corr}. Like for
luminescence, the general shape is reproduced very well in the
calculation. This good agreement lends support to the hypothesis that
PBE calculations yield very good vibrational frequencies, but
underestimate geometry relaxation.

Nevertheless, we see small discrepancies.  For examample, the
calculation shows a small peak at 2.11 eV, indicated by ``$\#$'' in
the plot. It originates from a localized phonon mode just above the
bulk phonon spectrum (less than 1 meV in our calculations), and this 
small peak seems to be absent in the experimental spectrum. 
[Interestingly, a localized phonon mode just above the bulk band has 
been observed in the infrared absorption spectrum for the ${\sE}\rightarrow{\sA1}$
transition~\cite{Kehayias2013} (cf.~Fig.~\ref{fig:nv}), even though,
clearly, different electronic states are involved in that transition.]
This localized mode is present in both PBE and HSE calculations (Fig.~\ref{fig:theory_abs}).
The energy of this localized mode is very sensitive to the parameters of
our calculations and its localized nature might be an artefact. 
Despite these two issues, the calculations show very good agreement
with experiment. In particular, if we look at the \textit{asymmetry}
of luminescence and absorption lineshapes in experiment, we see that
this asymmetry is reproduced very well.

We are now in the position to discuss why the luminescence lineshape,
presented in our previous work (Ref.~\onlinecite{Alkauskas2014}),
already showed good agreement with the experimental curve, in spite of
the fact that the methodology in that study was not as sophisticated
as what is presented in the current work.  The good agreement in
Ref.~\onlinecite{Alkauskas2014} was largely due to fortuitous
cancellation of two factors. (i) In Ref.~\onlinecite{Alkauskas2014}
only the coupling to symmetric modes $a_1$ was considered. As we now
know from Table~\ref{tab:HR_tot}, the contribution of $e$ modes to
luminescence is ${\sim} (16\mbox{--}18)\%$. Neglect of
$e$ modes should have led to an underestimation of the theoretical
Huang--Rhys factor compared to the experimental one. (ii) In
Ref.~\onlinecite{Alkauskas2014}, forces were calculated with HSE, but
vibrational frequencies of the defect system were calculated using
PBE, due to the high computational cost of HSE. Vibrational
frequencies in HSE are higher than in PBE, as we have seen in the
current work. Thus, using HSE forces with PBE frequencies in
Eqs.~\eqref{eq:partial} and~\eqref{eq:fc_displ} leads to an
overestimation of partial Huang--Rhys factors for coupling to
$a_1$ modes.  The two factors (i) and (ii) fortuitously compensated
each other to a large degree, leading to a good agreement of the
calculated lineshape with experiment.

\section{Conclusions}

\label{sec:conc}

In conclusion, we have presented a theoretical study of the
vibrational and vibronic structure of the negatively charged
nitrogen--vacancy center in diamond. Our main focus was the
calculation of luminescence and absorption lineshapes. We have
approached the dilute limit by embedding the NV center in supercells
of up to 64\,000 atoms. This resulted in converged spectral densities
of electron--phonon coupling throughout the whole phonon spectrum.  We
have developed a computationally tractable methodology to account for
the dynamical multi-mode Jahn--Teller effect, and have shown that the
use of this methodology is particularly important when studying
absorption at NV centers. Our calculations show that the vibrational
structure determined with the PBE functional agrees slightly better
with experiment than the one determined with the HSE
functional. Nevertheless, the geometry relaxation between the two
electronic states of the NV center is slightly underestimated in PBE,
and slightly overestimated in HSE (judging from the comparison of the
calculated total Huang--Rhys factor with the experimental one). This
indicates that, while being overall accurate, presently available
density functionals are still not ``perfect'' for describing subtle
features in optical lineshapes. We can obtain excellent agreement with
experiment for both luminescence and absorption by using PBE spectral
densities of electron--phonon coupling but scaling them with a factor
$\zeta=1.2$.

Looking forward, our work indicates that the continuing development of
more accurate density functionals, as well as computational advances
for a rigorous but tractable treatment of excited states will be an
essential feature of \textit{quantitative} first-principles
calculations for point defects. The methodology presented here advances first-principles
calculations of electron-phonon coupling \cite{giustino2017} for defects and will be
useful in the study and identification of other point defects in
solids. In particular, it is our hope that the methodology will help
in identifying and designing~\cite{Bassett2019} new quantum defects.

\section*{Acknowledgments}

This work has received funding from the European Union's Horizon 2020
research and innovation programme under grant agreement No.~820394
(project A{\sc steriqs}).
MD and NBM acknowledge funding from the Australian Research Council
(DE170100169 and DP170103098).
CGVdW was supported by the National Science Foundation (NSF) through
Enabling Quantum Leap: Convergent Accelerated Discovery Foundries for
Quantum Materials Science, Engineering and Information (Q-AMASE-i)
(DMR-1906325).
AA acknowledges the NSF Materials Research
Science and Engineering Centers (MRSEC) Program
(DMR-1720256) (Seed),
and NSF Q-AMASE-i (DMR-1906325) for funding visits to the University of California.
Computational resources were provided by the High Performance
Computing Center ``HPC Saul\.etekis'' in the Faculty of Physics,
Vilnius University, the Extreme Science and Engineering Discovery
Environment (XSEDE), which is supported by the NSF (ACI-1548562), and
the National Computational Infrastructure (NCI), which is supported by
the Australian Government.


\appendix

\section{Comparison of Jahn--Teller and Huang--Rhys treatment in emission}
\label{sec:apJT}

In Sec.~\ref{jt} we have shown that in the case of luminescence at
$T=0$ K the spectral functions $A_{e} (\hbar\omega)$ calculated by
means of the diagonalization of the vibronic Hamiltonian, on the one
hand, and based on a simpler Huang--Rhys approach, on the other, are
very close to each other (Fig.~\ref{fig:Ae-res-lum}). In this Appendix
we provide a rationale for this behavior and show that the agreement
is to a large degree accidental for the NV center.

Let us consider a single $e$ doublet with frequency $\omega$. In the
notation of Sec.~\ref{jt} the harmonic part of the Hamiltonian is
[cf.~Eq.~(\ref{eq:Hph})]
\begin{equation}
  \Hop_{\mathrm{0}} = \C{z} \sum_{\gamma\in x, y} \left ( -\frac{\hbar^2}{2}\frac{\partial^2}{\partial Q^2_{\gamma}}
    + \frac{1}{2}  \omega^2 Q_{\gamma}^2 \right ) .
\label{eq:Hph2}
\end{equation}
The linear Jahn--Teller interaction, Eq.~\eqref{eq:JT}, can be
rewritten employing the parameter $S_e$ introduced in Sec.~\ref{jt}
as:
\begin{equation}
  \Hop_{\mathrm{JT}} = \sqrt{2S_e\hbar\omega^3} \left ( \C{x}  Q_{x} + \C{y}  Q_{y} \right ) .
\label{eq:JT2}
\end{equation}
When the Jahn--Teller interaction is weak, $S_e \ll 1$, one can apply
first-order perturbation theory.  The ground-state eigenfunctions of
the unperturbed Hamiltonian, Eq.~(\ref{eq:Hph2}), are
$\ket{00;E_{\pm}} $ in the notation of Sec.~\ref{jt}. They correspond
to two angular momentum components $j=\pm\tfrac{1}{2}$. As the total
vibronic Hamiltonian for different values of $j$ is decoupled, we can
apply a simple non-degenerate perturbation theory to obtain
first-order corrections of ground-state vibronic wavefunctions due to
the presence of Jahn--Teller interactions Eq.~\eqref{eq:JT2}:
\begin{equation}
  \label{eq:JT_P1}
  \ket{\chi_0^{\mathrm{JT}};j=\pm\tfrac{1}{2}}
 = \mathcal{A}\left(\ket{00; E_{\pm}} -\sqrt{2S_e}\ket{1\pm 1;E_{\mp}}\right).
\end{equation}
Here $\mathcal{A}=1/\sqrt{1+2S_e}$ is the normalization factor.

\begin{figure}
\includegraphics{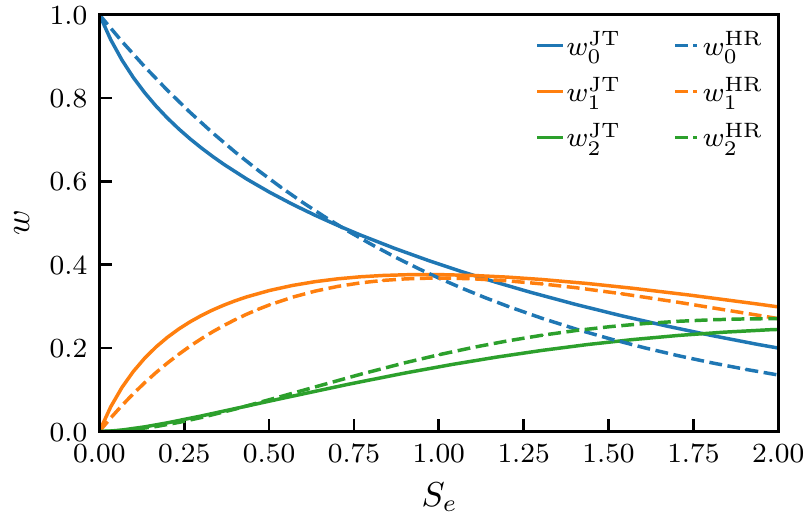}
\caption{Intensities of the zero-phonon line ($w_0$) and the first two
  phonon replicas ($w_1$ and $w_2$) in the JT (solid lines) and the HR
  (dashed lines) treatments as a function of the Huang--Rhys factor
  $S_e$.\label{fig:1mode_jt_vs_hr}}
\end{figure}

Our goal is to compare the JT treatment with the HR
treatment. Therefore, we need to find a Hamiltonian of the type
Eq.~\eqref{eq:HR_pert} that would yield the same Huang--Rhys factor
$S_e$ as above. There are many ways to achieve this; the final result
is independent of this choice. One particular realization is
\begin{equation}
\Hop_{\mathrm{HR}} = \C{z} \sqrt{S_e \hbar\omega^3}(Q_x + Q_y),
\end{equation}
to be compared with Eq.~\eqref{eq:JT2}. We will treat this as a
perturbation to the Hamiltonian Eq.~\eqref{eq:Hph2}.  The
linear-coupling of this form does not couple the two orbital states,
so we can only consider the vibrational part of the
wavefunction. Using the $\ket{n_{x}n_{y}}$ basis for our vibrational
states, we find that first-order perturbation theory gives
wavefunctions:
\begin{equation}
\label{eq:2}
\ket{\chi_0^{\mathrm{HR}}}= \mathcal{A}\left(\ket{00} - {\sqrt{\frac{S_e}{2}}}\ket{10} - {\sqrt{\frac{S_e}{2}}}\ket{01} \right).
\end{equation}
with the normalization factor $\mathcal{A}=1/\sqrt{1+S_e}$.

Using the above expressions we can now derive the intensity of the
first phonon replica (labeled $w_1$) in the normalized luminescence
spectrum.  When $S_e\ll 1$, we get:
\begin{eqnarray*}
  && w_1^{\mathrm{JT}} = \frac{2S_e}{1 + 2S_e}=2S_e + O(S^2_e)\\
  && w_1^{\mathrm{HR}} = \frac{S_e}{1 + S_e}= S_e + O(S^2_e).
\end{eqnarray*}
Thus, for the same value of the parameter $S_e$ the JT theory yields
the intensity of the first phonon peak twice as large as the HR
theory.

In Fig.~\ref{fig:1mode_jt_vs_hr} the analysis is extended for larger
values of $S_e$, where we compare the intensity of the ZPL $w_0$ as
well as of the first two phonon replicas in the two theories. In the
case of the HR theory an analytical result [Eq.~\eqref{eq:hr_overlap}]
was used, while numerical diagonalization of the vibronic Hamiltonian
was performed for the JT theory. For very small $S_e$, the numerical
results confirm the conclusions of the perturbation theory, i.e., that
the intensity of the first phonon replica is twice as large in the JT
treatment.  However, for $S_e\approx 0.5 - 1.0$, the values of $w_0$,
$w_1$, and $w_2$ are very close to each other.  For example, for
$S_e=0.75$ $w_0$ and $w_1$ differ by less than 4\%, while the values
of $w_2$ (which are smaller) differ by 15 \%.\@ The actual values of
$S_e$ for the NV center fall in this range (Table~\ref{tab:HR_tot}),
rationalizing the similarity of the HR and JT treatments for
luminescence. When $S_e$ is increased above this range, the HR theory
still performs rather well and can be considered as an approximation
to the JT treatment. However, one should be extremely cautious
applying the HR approach for JT systems with $S_e<0.5$, especially if
the focus is on the phonon sideband.  \vfill




\bibliography{references}

\end{document}


\title{Vibrational and vibronic structure of isolated point defects: the nitrogen-vacancy center in diamond. Supplemental material}

\author{Lukas Razinkovas}
\author{Marcus W. Doherty}
\author{Neil B. Manson}
\author{Chris G. Van de Walle}
\author{Audrius Alkauskas}

\noaffiliation
\maketitle	

\section{Decay of forces}
\label{sec:decay_forces}

\begin{figure}
\includegraphics[width=12cm]{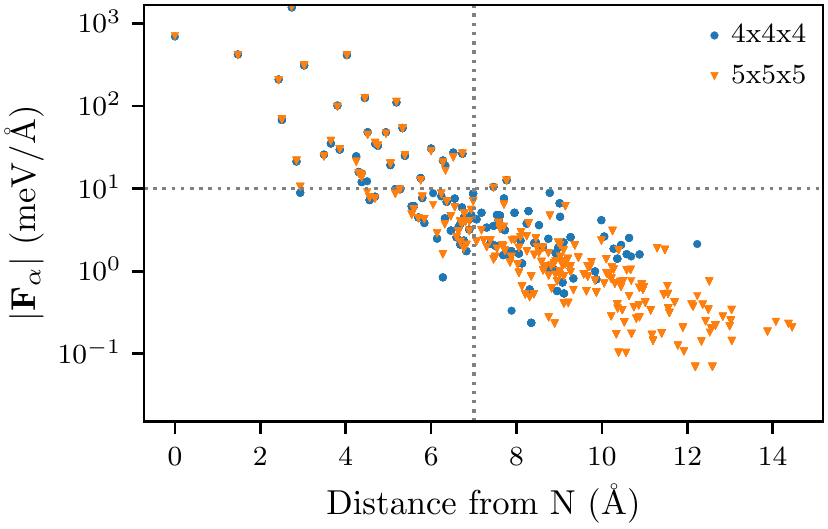}
\caption{Decay of forces on the atoms surround the NV center when the electronic state is changed from $\tE$ to $\tA2$ 
at the equilibrium geometry of the $\tE$ state. Results for two supercells are shown: $4\times 4\times 4$ (512 atomic sites)
and $5\times 5\times 5$ (1000 atomic sites)}
\label{fig:fconv}
\end{figure}

In Sec.~VI of the main text we discussed the rapid decay of forces on the atoms surrounding the NV center
when the electronic state is changed from $\tE$ to $\tA2$ at the equilibrium geometry of the $\tE$ state.
This is illustrated in Fig.~\ref{fig:fconv}, which shows the force on the atom as a function
of the distance from the nitrogen atom. In the figure we compare the results from two supercells: $4\times 4\times 4$ (512 atomic sites)
and $5\times 5\times 5$ (1000 atomic sites). The largest forces are experienced by the carbon atoms around the vacant site and are slightly 
above 1000 meV/\AA. However, forces quickly decay away from the NV center. At distances $d>7$ \AA$ $ from the nitrogen atom the forces 
become smaller than 10 meV/\AA. Importantly, we find that for atoms that are both closer than $d=7$ \AA$ $ to the N site and those 
with forces $F>10$ meV/\AA$ $ (upper left quadrant), the results from the two supercells fall on top of each other. This indicates 
that these forces are already converged as a function of the supercell size. For atoms at distances $d<7$ \AA$ $,
but with forces $F<10$ meV/\AA$ $ (lower left quadrant) the difference in forces is slightly larger, but the overall agreement is 
still very good. Supercell effects are much larger for atoms further than $d=7$ \AA$ $ from the N atom, however, the forces on the atoms
are very small. In our calculations we set all these forces to zero when calculating spectral densities.
We estimate that the error associated with this approximation for calculated spectral densities and the total 
Huang-Rhys factor is $<1$ \%.

The same reasoning as above holds for forces when the state is changed from $\tA2$ to $\tE$ at the equilibrium geometry
of the $\tA2$ state.

\section{Choice of frequencies in the calculation of luminescence and absorption lineshapes}
\label{sec:pma}

When calculating absorption and emission lineshapes in Secs.~VI,
VII, and VIII of the main text, we employed the equal-mode
approximation, whereby the vibrational modes and frequencies in the
ground and excited states were taken to be identical. In this Section
we discuss the validity of this simplification. We stress at the
outset that the validity of the approximation can be rigorously tested
only in systems with a small number of vibrational degrees of freedom,
where the exact calculation is possible (see, e.g., Fig.~2 in
Ref.~\onlinecite{alkauskas2012first}). Here we resort to a qualitative
analysis.

\begin{figure}
\includegraphics[width=9cm]{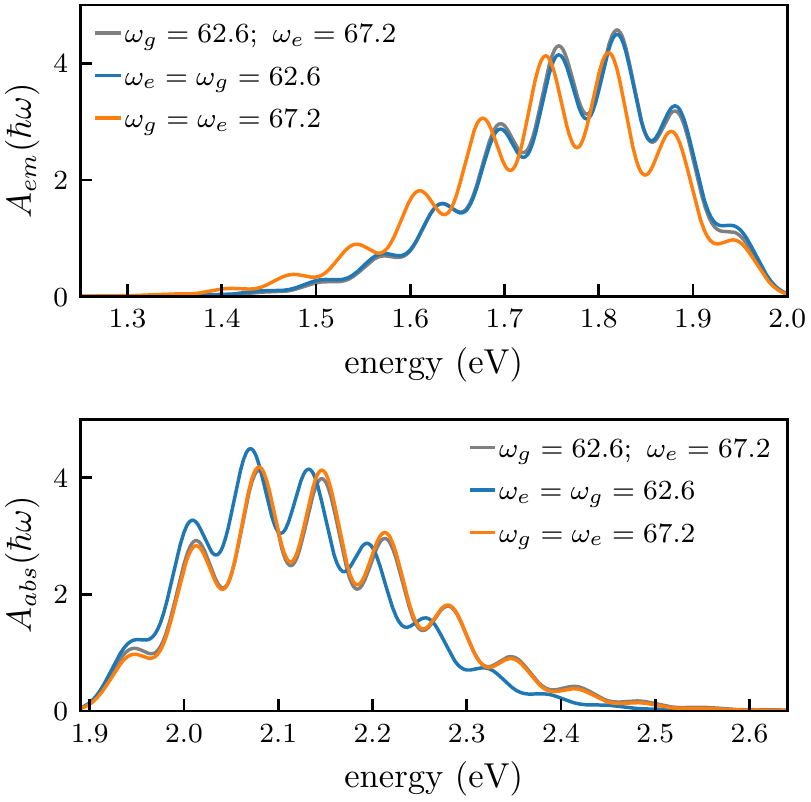}
\caption{Spectral functions (in 1/eV) for luminescence (a) and absorption (b)
  in the 1D approximation.
  Gray: using actual frequencies in the ground and the excited state;
  blue: using the frequency of the ground state;
  orange: using the frequency of the excited state.
  The ZPL energy was set to the experimental value, $E_{\text{ZPL}} = 1.945$
  eV, and a smearing parameter $\sigma = 25$ meV was used.}
\label{fig:ema}
\end{figure}

The difference between the two sets of modes can be quantified by
calculating the average frequency of modes that contribute to the
optical lineshapes. This average is defined as:
\begin{equation}
\Omega_{\{e,g\}}^2 = \frac{\sum_k \omega_{{\{e,g\}};k}^2 \Delta Q_k^2}{\sum_k \Delta Q_k^2}.
\end{equation}
The sum runs over all vibrational modes of $a_1$ and $e$
symmetry. $\Delta Q_k$ is defined in Eq.~(31) of the main text for $a_1$ modes
and an analogous set of equations Eqs.~(45) and
(46) of the main text for $e$ modes. We obtain $\hbar\Omega_{g}=62.6$ meV and
$\hbar\Omega_{e}=67.2$ meV for average frequencies in the ground and
the excited state. This result indicates that the frequencies of modes
that contribute to optical lineshapes are on average higher in the
excited state than in the ground state.

Having determined average frequencies we can now calculate
luminescence and absorption lineshapes in a model system of
one-dimensional quantum harmonic oscillators described by these
frequencies (see Ref.~\onlinecite{alkauskas2012first} for more
details).  In Fig.~\ref{fig:ema} we compare the lineshapes calculated
using the actual frequencies that are different for ground and excited
states, with two approximations: in the first approximation the two
frequencies are taken to coincide with that of the ground state
$\hbar\Omega_{g}=62.6$ meV (orange lines), while in the second
approximation they coincide with that of the excited state
$\hbar\Omega_{e}=67.2$ meV (green lines).  The results show that the
``true'' lineshape is approximated extremely well by setting the
frequency to that of the {\it final} state: {\it i.e.}, the ground
state in the case of luminescence, and the excited state in the case
of absorption.  In contrast, setting the frequency to that of the
initial state does a poor job.

\section{Benchmarking the diagonalization of the Jahn-Teller Hamiltonian}
\label{sec:bench_diag}

\begin{figure}
\includegraphics{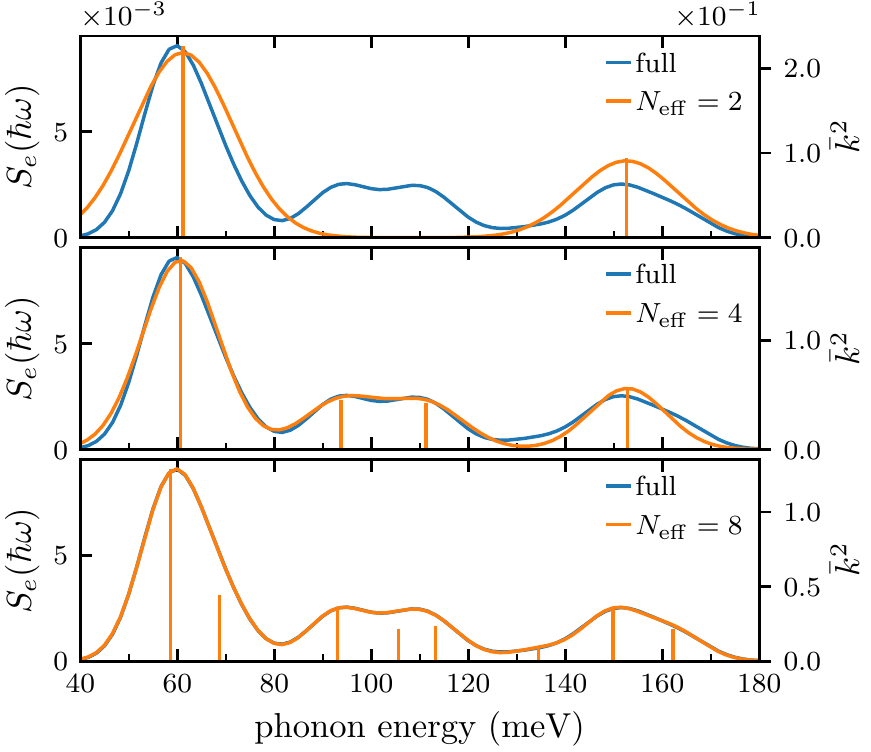}
\caption{ Convergence of
  $S^{(\mathrm{eff})}_{e}(\hbar\omega)$  towards $S_{e}(\hbar\omega)$
  [Eq.~(49) of the main text] when increasing the number of effective modes
  $N_{\mathrm{eff}}$. The results are for the NV center in the
  $2\times 2\times 2$ supercell.
  Spectral densities in units 1/meV.
  }
\label{fig:2x2x2a}
\end{figure}

\begin{figure}
\includegraphics[width=8.5cm]{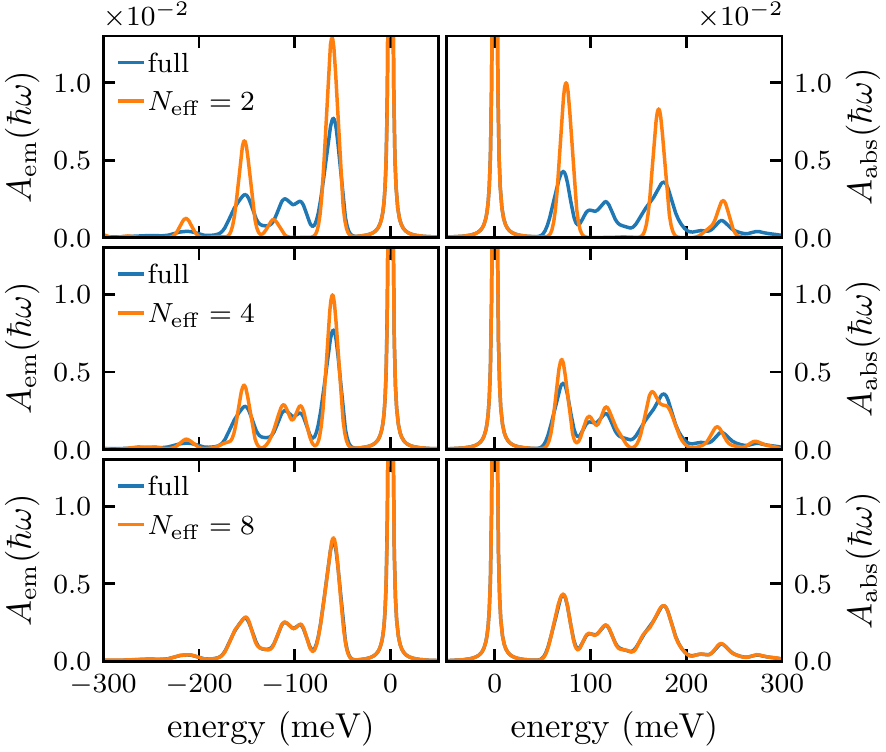}
\caption{Convergence of the spectral function
  $A_e(\hbar\omega)$ [in units 1/meV] for emission (left) and absorption (right) as a
  function of $N_{\mathrm{eff}}$, the number of effective $e$ modes
  included in the calculation.  The results are for the NV center in
  the $2\times 2\times 2$ fcc supercell.}
\label{fig:fitting2x2x2b}
\end{figure}

In this Section we illustrate the procedure of the diagonalization of
the Jahn-Teller Hamiltonian outlined in Sec.~VII D of the main text in
the case of the $2\times 2\times 2$ fcc supercell, for which it is
possible to diagonalize the vibronic Hamiltonian including all 62 $e$
doublets. In Fig.~\ref{fig:2x2x2a} we show the approximation of
$S_e(\hbar\omega)$ with $S^{(\mathrm{eff})}_{e}(\hbar\omega)$ for
$N_{\mathrm{eff}} =2, 4, 8$ [Eq.~(49) of the main text]. We see that in the
case of $N_{\mathrm{eff}} =8$ the approximate spectral density
$S^{(\mathrm{eff})}_{e}(\hbar\omega)$ is indistinguishable from the
real spectral density $S_e(\hbar\omega)$.

For the diagonalization of the Jahn-Teller Hamiltonian we use
$n_{\text{tot}}=3$.  Although this basis is not converged with respect
to $n_{\text{tot}}$, it is sufficient for benchmarking purposes, as we
keep $n_{\text{tot}}$ the same for all calculations with effective
modes.  The resulting spectral functions $A_e(\hbar \omega)$ for
absorption and emission, calculated from
$S^{(\mathrm{eff})}_{e}(\hbar\omega)$, are compared to the ones
calculated from $S_{e}(\hbar\omega)$ that include {\it all} vibrations
for the given supercell in Fig.~\ref{fig:fitting2x2x2b}.  We see that
the result with $N_{\mathrm{eff}}=8$ reproduces the full result very
accurately.  Note that the spectral function $S_{e}(\hbar\omega)$ for
the $2\times 2\times 2$ supercell is not yet converged with respect to
the supercell size. The purpose of Fig.~\ref{fig:fitting2x2x2b} is to
illustrate our methodology for the largest supercell for which the
full calculation is still possible.

\bibliography{../references}